\documentclass[aps,prb,reprint,showpacs,superscriptaddress]{revtex4-1}
\usepackage{graphicx}
\usepackage{color} 
\usepackage{longtable}
\begin{document}


\title{Effect of Rotation on Elastic Moduli of Solid $^4$He}



\author{T. Tsuiki}
\affiliation{Department of Physics, Keio University, Yokohama 223-8552, Japan}
\affiliation{RIKEN CEMS, Wako 351-0198, Japan}
\author{D. Takahashi}
\affiliation{Center for Liberal Arts and Sciences, Ashikaga Institute of Technology, Ashikaga 326-8558, Japan}
\affiliation{RIKEN CEMS, Wako 351-0198, Japan}
\author{S. Murakawa}
\affiliation{Cryogenic Research Center, University of Tokyo, Bunkyo-ku, Tokyo 113-0032, Japan}
\affiliation{RIKEN CEMS, Wako 351-0198, Japan}
\author{Y. Okuda}
\affiliation{RIKEN CEMS, Wako 351-0198, Japan}
\author{K. Kono}
\affiliation{RIKEN CEMS, Wako 351-0198, Japan}
\author{K. Shirahama}
\affiliation{Department of Physics, Keio University, Yokohama 223-8552, Japan}
\affiliation{RIKEN CEMS, Wako 351-0198, Japan}


\date{\today}

\begin{abstract}
We report measurements of elastic moduli of hcp solid $^4$He down to 15 mK when the samples are rotated unidirectionally. 
Recent investigations have revealed that the elastic behavior of solid $^4$He is dominated by gliding of dislocations and pinning of them by $^3$He impurities, which move in the solid like Bloch waves (impuritons). 
Motivated by the recent controversy of torsional oscillator studies, we have preformed direct measurements of shear and Young's moduli of annular solid $^4$He using pairs of quarter-circle shape piezoelectric transducers (PZTs) while the whole apparatus is rotated with angular velocity $\Omega$ up to 4 rad/s. 
We have found that shear modulus $\mu$ is suppressed by rotation below 80 mK, when shear strain applied by PZT exceeds a critical value, 
above which $\mu$ decreases because the shear strain unbinds dislocations from $^3$He impurities.
The rotation - induced decrement of $\mu$ at $\Omega = 4$ rad/s is about 14.7(12.3) $\%$ of the total change of temperature dependent $\mu$ for solid samples of pressure 3.6(5.4) MPa.
The decrements indicate that the probability of pinning of $^3$He on dislocation segment, $G$, decreases by several orders of magnitude. 
We propose that the motion of $^3$He impuritons under rotation becomes strongly anisotropic by the Coriolis force, resulting a decrease in $G$ for dislocation lines aligning parallel to the rotation axis.
\end{abstract}


\maketitle

\section{Introduction\label{Introduction}}
Solid $^4$He, a unique bosonic quantum solid, shows various quantum effects\cite{Andreev,Pushkarov,Varma}.
$^4$He atoms exchange frequently their positions between the lattice sites. 
The exchange leads to a dramatic effect in the motion of isotopic $^3$He impurity, which is inevitably contained in any $^4$He samples.
A $^3$He impurity can move as a Bloch wave in the periodic potential formed by surrounding $^4$He atoms, a situation like electrons in crystal. 
Such a Bloch state is called \textit{impuriton} or mass fluctuation waves\cite{Andreev,Pushkarov,guyer1970mass,guyer1971excitations,richards1972evidence,huang1975quantum,esel1978quantum}. 
Since there are particle exchanges, bosonic solid $^4$He may exhibit superfluidity by Bose - Einstein condensation while its crystalline structure is kept\cite{andreev1969quantum,chester1970speculations,leggett1970can}.
Search for the coexistence of superfluid and crystalline orders, called supersolidity, was sparked by the discovery of torsional oscillator (TO) anomaly in 2004 by Kim and Chan\cite{kim2004probable,kim2004observation}. 
It is nowadays realized that the TO behavior found by Kim and Chan, a drop of resonant period associated with dissipation, turned out to be originated from the elastic change of solid $^4$He.
Even so, supersolid state of matter has been one of the most fundamental problems in condensed matter physics and bosonic $^4$He is still a candidate to realize such a intriguing state \cite{nyeki2017intertwined}.

Intensive studies of the putative supersolid state have revealed anomalous behaviors in elastic properties of solid $^4$He at subkelvin temperatures\cite{Day_and_Beamish_Nature_2007,Day_etal_PRB_2009,Day_Syshchenko_and_Beamish_PRL_2010,Beamish_etal_PRB_2012,Haziot_etal_PRB_2013}. 
The shear modulus of solid $^4$He, $\mu$, shows a minimum at temperatures $200 < T < 500$ mK. 
Below 200 mK, it increases with decresing $T$ and saturates below about 50 mK. 
The anomalous increase in shear modulus below 0.2 K has been attributed to pinning of dislocation network existing in solid $^4$He by $^3$He impurity atoms\cite{Day_and_Beamish_Nature_2007,Iwasa_JLTP_2013}. 
The anomalous elastic behavior owing to the dislocation motion and its pinning by $^3$He is a consequence of quantum nature in solid $^4$He.
First, quantum fluctuation of $^4$He atoms around lattice sites makes the Peierls potential so small that each dislocation segment between intersections can easily vibrate like a string in which both ends are fixed.
Second, $^3$He dilutely dispersed in solid $^4$He form a Bloch state with energy bands and hence can move as an impuriton which has a group velocity and a band effective mass determined by band structure, especially by bandwidth. 
The quantum motion of impurity $^3$He was studied in 1970-80's\cite{guyer1970mass,guyer1971excitations,richards1972evidence,huang1975quantum,esel1978quantum}. 
Magnetic resonance experiments and theoretical studies revealed that at low enough concentration $^3$He impurity behaves as impuritons,
but the interaction between dislocations and impuritons was not discussed. 
In real solid $^4$He, dislocation networks are strongly pinned at their intersections (nodes) so that they cannot move as a whole but only the segments between nodes oscillate in particular gliding directions. 
It should be emphasized that, $^3$He needs to move freely in solid $^4$He to realize the pinning of dislocations and consequently the increase in shear modulus, and the movement of $^3$He near 0 K can be realized only by the impuriton mechanism.

In this paper, we report on new phenomenon related to the elasticity of solid $^4$He below 0.1 K, a decrement of shear modulus when solid sample is steadily rotated.
We have observed decrements in two solid samples with different pressures. 
The decrements suggest that pinning of dislocation network by $^3$He atoms is disturbed by rotation. 
We propose a possible mechanism for the suppression of $^3$He pinning by considering the quantum nature of solid and impurity: 
The Coriolis force acting on $^3$He impuritons makes their motion strongly anisotropic. 
This anisotropy of $^3$He diffusion can decrease the probability of attaching $^3$He to dislocations.

One of our motivation for the present work is to solve the controversy in the TO measurements under rotation. 
Choi \textit{et al}. found that the apparent supersolid fraction, i.e. the magnitude of period drop, obtained from a TO containing annulus solid $^4$He sample is strongly suppressed by steady DC rotation\cite{Choi_and_Takahashi_etal_Science_2010,Choi_and_Takahashi_etal_PRL_2012,Choi_etal_PRB_2012}.
They attributed the suppression of the period drop to the intrusion of quantum vortices in supersolid.
As the TO period shift is not originated from supersolidity, the origin of the rotation effect discovered by Choi \textit{et al}. is now a most important puzzle:
It has been established by theoretical considerations and by Finite-Element Method simulations that increase in shear modulus of solid $^4$He contained in the bob of a TO stiffens the TO bob and thus decrease the TO period. 
This stiffening - induced period drop was mistaken as a superfluid response of solid He inside.
Given that the TO period drop is entirely due to elastic change in solid $^4$He, shear modulus has to be influenced by rotation.
Choi \textit{et al}. measured shear modulus under rotation by a pair of piezoelectric transducers (PZTs) located in the center of their TO, 
but they did not observe rotation effect\cite{Choi_and_Takahashi_etal_Science_2010}. 
In our present setup, we measure shear modulus in the direction of circumference of annular solid samples so that it is possible to conclude the origin of rotation effect observed by Choi \textit{et al}..

We organize this paper as follows: 
In the following Section, we describe elastic properties solid $^4$He and currently argued interpretation. 
In Sec. \ref{Experimental}, we show the experimental method to measure shear and Young’s modulus for annulus solid $^4$He sample under rotation. 
In Sec. \ref{Results}, experimental results with and without rotation are shown, followed by the interpretation for these results in terms of impuriton model in Sec. \ref{Interpretation}. 
In Sec. \ref{Discussion} we give some discussion and the paper is summarized in Sec. \ref{Conclusion}.

\section{Backgrounds\label{Backgrounds}}
\subsection{Shear modulus anomaly and dislocation in solid $^4$He}
Since 1970's, it has been well established that solid helium contains dislocations as a defect by measurements of plastic deformation\cite{suzuki1973plastic}, ultrasound\cite{wanner1976evidence,iwasa1980sound} and X-ray\cite{iwasa1987observation}. 
As in dislocations in ordinary matters such as metals, dislocations in helium compose a network structure.
The nodes of dislocation network are strongly fixed, while dislocation segments between nodes can vibrate like strings in response to stress\cite{granato1956theory}. 
Ultrasonic studies showed that $^3$He impurity atoms attach to dislocation segments at low temperatures and disturb their oscillations responding to ultrasound\cite{iwasa1980sound}. 
It is suggested that, in hcp solid $^4$He, $^3$He impurity atoms are trapped to edge and mixed dislocations in which both the dislocation lines and their Burgers vectors are aligned to the basal plane of hcp lattice ((0001) plane). 
This is because elastic deformation around the core of dislocations forms an attractive potential on the side where atomic density is low.
Note that another side has higher atomic density, and exerts a repulsive force.
Screw dislocations do not attract $^3$He.

The elastic modulus of solid $^4$He at subkelvin temperatures is determined by the motion of dislocation segments\cite{Day_and_Beamish_Nature_2007,Day_etal_PRB_2009,Day_Syshchenko_and_Beamish_PRL_2010,Beamish_etal_PRB_2012,zhou2012dislocation,haziot2013giant,Haziot_etal_PRB_2013,Iwasa_JLTP_2013}. 
As solid is cooled below 1 K, dislocation movement is damped by the collision with phonons so that the shear modulus is relatively high. 
As temperature decreases, the number of phonons decreases 
and dislocation segments become free to glide in the basel plane in response to applied shear, while the intersections act as nodes and are immobile.
The shear modulus has a minimum at temperatures 0.2 - 0.5 K, depending on $^3$He concentration.
At lower temperatures (e.g. below 0.2 K), shear modulus starts to increase because $^3$He atoms start to stick to dislocation segments and disturb their motion.
Shear modulus gradually approaches to its intrinsic value below 50 mK. 

\begin{figure}
\includegraphics[width=0.35\textwidth]{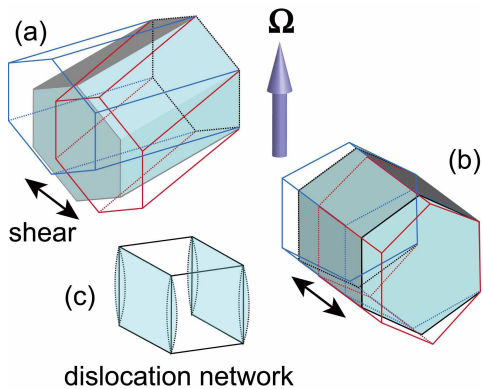}
\caption{(Color online) Two cases of shear motions of hcp $^4$He that contribute to change in $c_{44}$. (a) The hcp basal plane glides to [1010], [1100] or [0110] directions. (b) (1010), (1100) or (0110) planes glide to the direction of $c$ ([0001]) axis. (c) Schematic drawing of dislocation network (cubic shape is assumed for simplicity). The vibration of dislocation lines perpendicular to glide direction and in parallel to the gliding plane (shown by dashed lines) only contributes to shear stress. 
The direction of sample rotation $\Omega$ is also shown by an arrow for later discussion. 
\label{fig:HCP_Dislocation}}
\end{figure}

The temperature and $^3$He concentration dependencies of shear modulus were first measured by Day and Beamish for polycrystalline solid samples\cite{Day_and_Beamish_Nature_2007}.
Later studies by Balibar and coworkers in which the orientations of single crystals to PZTs were precisely controlled revealed that only the elastic constant $c_{44}$ is responsible for the increase in shear modulus below 0.2 K\cite{haziot2013giant}. 
The situation is shown in Fig. \ref{fig:HCP_Dislocation}.
$c_{44}$ corresponds to the elastic tensor component when the hcp basal plane glides to [1010], [1100] or [0110] directions (Fig. \ref{fig:HCP_Dislocation}(a)),
and when (1010), (1100) or (0110) planes glide to the direction of $c$ ([0001]) axis ((b)). 
The dislocations which mainly determine $c_{44}$ are aligned perpendicularly to shear strain and in parallel to the gliding plane, illustrated as vertical lines in Fig. \ref{fig:HCP_Dislocation}(c).
The vibration of the vertical dislocations glide the planes and can relax shear stress.
It is remarkable in hcp $^4$He that the glide motion occurs at very low applied shear strain $\epsilon \sim 10^{-9}$. 

Although the microscopic mechanism of trapping $^3$He to dislocation core has not been understood,
the effect of $^3$He impurity on shear modulus has been phenomenologically discussed and has successfully explained the dependencies of shear modulus on temperature and applied shear stress\cite{fefferman2014dislocation}. 
We will discuss the detail of the models in Sec. \ref{Interpretation}.

\subsection{Impuriton: quantum impurity $^3$He in solid $^4$He}
Elastic anomaly below 0.2 K occurs even when imposed shear stress is so small that the displacement of dislocation  is less than 10 nm.
Since the average length of dislocation segments is on the order of 10 $\mu$m and should be comparable to the average separation of segments, 
$^3$He atoms move a long distance in solid $^4$He, in order to be trapped by dislocation core.
Definitely, such mobility of $^3$He near 0 K can be realized only by quantum effects.

$^3$He impurities are delocalized and behave as quasiparticles like Bloch waves in periodic potential of the $^4$He matrix, 
hence the wavenumber $k$ of the quasiparticles become a good quantum number and the motion is determined by energy bands.
The quasiparticle was called \textit{impuriton} by Richards\cite{richards1972evidence} or \textit{mass fluctuation waves} by Guyer\cite{guyer1970mass,guyer1971excitations}. 
The impuriton picture for $^3$He in solid $^4$He was theoretically proposed by Andreev and Lifshitz\cite{andreev1969quantum}, 
in which they called it \textit{defecton}, and by the above authors, 
and then experimentally confirmed by many NMR studies for $^3$He concentration from 0.02 (2 \% molar ratio of $^3$He) down to $10^{-5}$ (10 ppm)\cite{greenberg1971isotopic,richards1972evidence,esel1978quantum}.
Note that the impuriton picture becomes better as the concentration decreases.
Therefore, although no NMR experiments were performed for commercial $^4$He gas in which typical $^3$He concentration is less than 1 ppm, 
one may expect that $^3$He behaves as impuriton in such a low concentration sample.

The most important feature of $^3$He impuriton is that the energy bandwidth is extremely small and it strongly depends on the molar volume (i.e. pressure) of solid $^4$He. 
Exchange energy between $^3$He and $^4$He, $J_{34}$, at the highest molar volume (near the melting pressure) 21 cm$^3$/mole is of the order of $10^{-5}$ K, which corresponds to 1 MHz in frequency\cite{Andreev,Pushkarov}. 
The corresponding energy bandwidth $\Delta$ is determined by crystal structure and is about $10J_{34}$ for hcp lattice, hence of the order of $10^{-4}$ K, the temperature of which is far smaller than the temperature we perform experiment. 
With increasing pressure (decreasing molar volume), the bandwidth rapidly decreases:
At pressures of the solid samples in the present work, $P = 3.6$ and 5.4 MPa, the bandwidth are less than 100 kHz and 10 kHz, respectively. 
Accompanied with this small bandwidth, the group velocity $v_{\mathrm{g}}$ of impuritons also becomes small: 
As shown in Sec. \ref{Interpretation}, $v_{\mathrm{g}}$ is of the order of 1 mm/s near melting, but it decreases down to 100 nm/s at $P = 5.4$ MPa.
Correspondingly, the effective mass of impuriton, $m^*$, is extremely large:
$m^*$ is $10^4 m_3$ where $m_3$ is the bare mass of a $^3$He atom ($7 \times 10^{-27}$ kg) at the melting pressure, and 
it exceeds $10^6 m_3$ at 5.4 MPa. 
As discussed in Sec. \ref{Discussion}, the large effective mass and small group velocity of impuritons are essential for the rotation effect on shear modulus.

The validity of the impuriton model at high pressures (e.g. $3 \sim 15$ MPa) is an important problem because many TO experiments have been done at such high pressures. 
According to the NMR experiment performed by Greenberg \textit{et al}., the exchange frequency of impuritons $J_{34}$ was measured up to 4 MPa (20 cc/mole), and it was found to be less than 1 kHz at the largest pressure\cite{greenberg1971isotopic}. 
Therefore, the impuriton model is valid at pressures up to 4 MPa.
The pressure of one of our solid samples is 3.6 MPa, which is well below this pressure.
Whether the impuriton model correctly describe the behavior of $^3$He at pressures higher than 4 MPa is yet to be confirmed experimentally. 
In a TO measurement by Kim and Chan, it was found that the apparent supersolid response was observed for many solid samples up to 15 MPa, 
where the onset temperature of the apparent supersolidity showed \textit{no} pressure dependence\cite{kim2006supersolid}. 
If the supersolid - like behaviors observed in the high pressure samples are entirely caused by the elastic anomaly of solid $^4$He,
$^3$He impurities have to \textit{run a macroscopic distance} even in such high pressure samples, 
otherwise the $^3$He - dislocation pinning model would never work out. 
There is also possibility that $^3$He can hop site to site by quantum tunneling without forming impuriton band.
It is therefore concluded that the picture of "dislocation pinning by $^3$He moving quantum mechanically" is valid for solid at pressures up to 15 MPa.

\subsection{Effect of rotation on properties of solid $^4$He: previous studies}
Far before the 2004 Kim - Chan experiment, Pushkarov theoretically studied the effect of rotation on defecton quasiparticles, in which he focused on zero point vacancy in solid $^4$He\cite{Pushkarov,pushkarov2001quasiparticle,pushkarov2012vacancy}. 
He discussed the motion of quasiparticles under sample rotation in terms of a Fokker - Planck equation, 
and found that quasiparticle diffusion is strongly suppressed in the direction perpendicular to the rotation axis\cite{Pushkarov,pushkarov2001quasiparticle}. 
When crystal is steadily rotated, the temperature dependence of diffusion coefficient perpendicular to the rotation axis change dramatically from $D \propto T^{-7}$, which is the behavior under no rotation and is determined by phonon scattering, to $T^9$ dependence at subkelvin temperatures, while the diffusion constant parallel to the rotation axis remains unchanged. 
This dramatic anisotropy of defecton diffusion is essentially the same conclusion as our consideration focussing on circular motion of impuritons by Coriolis force, which is discussed in Secs. \ref{Interpretation} and \ref{Discussion}. 
Pushkarov also claimed that the centrifugal force tends to change the distribution of defectons in the radial direction of a cylindrical solid\cite{pushkarov2012vacancy}. 
It is essentially an ordinary phenomenon of centrifuge, and we find that it is negligibly small for rotation speed of the order of 1 rad/s, even compared to the effect of gravity on the vertical distribution of $^3$He impuritons.

In order to study the possible quantized vortices in supersolid, TO experiments have been performed using several rotatable dilution refrigerators in the world.
Choi \textit{et al}. employed a TO containing annular solid $^4$He and a rotating dilution refrigerator at RIKEN, which is also used by us in the present work\cite{Choi_and_Takahashi_etal_Science_2010,Choi_and_Takahashi_etal_PRL_2012,Choi_etal_PRB_2012}. 
They observed that the magnitude of period drop, which was previously called "non-classical rotational inertia fraction" (NCRIF), 
is suppressed when unidirectional rotation is imposed to the torsional oscillation. 
As the rotation speed increases, the magnitude of the period drop decreases and the accompanying dissipation increases\cite{Choi_and_Takahashi_etal_Science_2010,Choi_and_Takahashi_etal_PRL_2012}. 
The oscillator used by Choi \textit{et al}. had a PZT pair at the center of the torsional bob so that shear modulus of the solid other than the annular part was measured independently. 
In contrast to the TO period, no significant change in shear modulus was observed when the apparatus was rotated\cite{Choi_and_Takahashi_etal_Science_2010,Choi_etal_PRB_2012}. 
It was thus concluded that the rotation effect on TO frequency is related to supersolidity. 
They further studied TO response by changing rotation speed continuously and observed a step - like TO period change, which was again interpreted as a vortex intrusion\cite{Choi_and_Takahashi_etal_PRL_2012}. 
Since any TO responses were revealed not to be originated from supersolidity, these experimental results of Choi \textit{et al} should be reconsidered on the basis of elastic behavior of solid $^4$He.

Other TO experiments under rotation performed by Yagi \textit{et al}.\cite{yagi2011probable} and Fear \textit{et al}.\cite{fear2016no} employed rotating cryostats of ISSP, Univ. Tokyo and of Univ. Manchester, respectively. 
They claimed that no significant rotation effect was observed in the period of TO containing cylindrical or annular solid $^4$He samples. 
Especially, Fear \textit{et al}.\cite{fear2016no} employed annular solid samples, which were similar to that used by Choi \textit{et al}., but they did not observe any rotation effect.
Fear \textit{et al}. claimed that the rotation effect observed by Choi \textit{et al}. should be originated from mechanical noise caused by rotating cryostat. 

This experimental controversy has recently been studied by two TO experiments using the rotating cryostat at RIKEN. 
One experiment done by Jaewon Choi \textit{et al}. employed a TO which has two annular tubes (donut shape) connected with a torsion rod\cite{choi2017superfluid}. 
This oscillator was specially arranged to make elastic contribution to the resonant period as small as possible (by no use of epoxy, and so on), 
and to have multiple resonant frequencies to detect the frequency - independent component in the period change, which can be originated from supersolidity\cite{choi2015frequency}.
The change in period drop by rotation up to 4 rad/s was small and did not reproduce the result of Choi \textit{et al}.\cite{choi2017superfluid}. 
Another experiment based on completely opposite idea by Tsuiki \textit{et al}. used a multiple - frequency TO which is very \textit{sensitive} to elasticity of solid $^4$He inside the bob\cite{TsuikiPreparation}.
This TO was firstly proposed by Reppy \textit{et al.} and is called a floating - core TO\cite{reppy2012interpreting}, in which the annular solid sample is formed between the outer shell and the inner core, which is hung by a torsion rod from the outer shell. 
In this elasticity - sensitive TO study, suppressions of TO period drop have been observed at intermediate temperatures depending on TO amplitude (i.e. applied shear stress), but the suppression at lowest temperatures is much smaller than that observed by Choi \textit{et al}.\cite{TsuikiPreparation}. 
Since both the experiments by Jaewon Choi \textit{et al}. and Tsuiki \textit{et al}. were done with using the same rotating cryostat at RIKEN\cite{takahashi2006new}, 
the possibility of the spurious effect by mechanical noise, which was claimed by Fear \textit{et al.}\cite{fear2016no}, is denied. 
The large rotation effect observed by Choi \textit{et al} is yet to be understood.
In Sec. \ref{Discussion} we will give a possible interpretation for the discrepancy in TO studies.

\section{Experimental setup\label{Experimental}}
\begin{figure}
\includegraphics[width=0.35\textwidth]{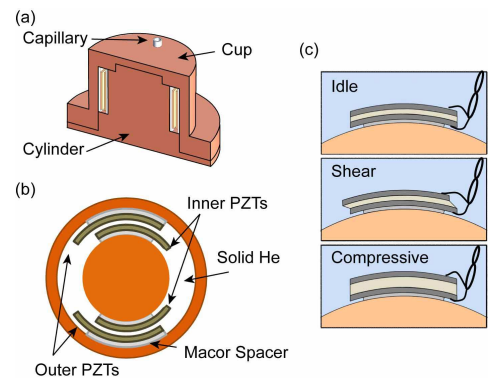}
\caption{(Color online) The experimental cell. (a) Schematic view. Capillary is used for supplying line of $^4$He. (b) Inner structure. The gaps between the PZTs are 0.5 mm. (c) Movements of two PZT's. Top : neutral position. The middle and the bottom express movements of shear and compressive PZTs, respectively.\label{fig:elasticity}}
\end{figure}

Our sample cell to measure elastic moduli is schematically shown in Fig. \ref{fig:elasticity}(a) and (b).
A thin annular channel is formed by gluing a cylinder to a container, both of which are made of BeCu alloy.
The inner and outer radii, and the height of the channel are 8.0, 10.5, and 12.0 mm, respectively.
Two pairs of quarter - circle shape PZTs (10 mm height and 0.5 mm thick)\cite{FUJI} are glued to the inner and outer walls with inserting round shaped Macor plates for insulation.
The arc lengths are 14.14 and 12.73 mm for outer and inner PZT, respectively.
The gaps between the inner and outer PZTs, i.e. the thickness of solid He sample, are set to 0.5 mm.

We show in Fig. \ref{fig:elasticity}(c) movements of two PZTs.
When a voltage is applied, one of the PZTs moves in the azimuthal direction, as shown in the middle. 
This motion gives a shear to the solid He sample between the pair PZT, so the PZT pair is used to measure shear modulus in the \textit{azimuthal} direction of the annular solid, i.e. the same direction as the rotation - induced velocity.
Another PZT expands in the radial direction as shown in the bottom of the cartoon, hence the PZT compresses the solid sample. 
This PZT pair applies a strain to the solid and detect the resultant stress in the \textit{radial} direction. 
It therefore measures Young's modulus.
We refer to the former PZT pair as the "shear" PZT, whereas the latter as the "compressive" PZT.

With these PZTs, we measure elasticity of solid $^4$He as follows:
We applied an AC driving voltage $V_{\mathrm {AC}}$ with frequency $f = 1.0$ kHz to the inner PZT which acts as drive PZT. 
The PZT produces an AC strain $\epsilon = V_{\mathrm {AC}} d_{\mathrm {PZT}} / D$ to solid $^4$He, where $d_{\mathrm {PZT}}$ is the piezoelectric constant, and $D$ the gap between the drive and detection PZTs, 0.5 mm.
We estimate $d_{\mathrm {PZT}}$ at low temperature (below 0.5 K) to be $7.91 (7.35) \times 10^{-11}$ m/V for the shear (compressive) PZT, respectively.
The strain causes stress $\sigma = \lambda \epsilon$. 
The elastic modulus of solid $^4$He $\lambda$ (the shear modulus $\mu$ or the Young's modulus $E$) is evaluated by measuring the piezoelectric current $I$ induced on the outer detection-PZT as $\lambda = (D/2\pi A d_{\mathrm {PZT}}^2) (I/fV_{\mathrm {AC}})$, where $A$ is overlapping area of drive and detection PZTs.
Since the current $I$ is typically an order of 1 pA, a current preamplifier with a fixed gain of $5 \times 10^8$ V/A and lock-in technique are used to measure the current.
The effect of electrical crosstalks between the drive and detection PZTs, which affects the phase difference between $I$ and $V_{\mathrm {AC}}$, is carefully removed using the crosstalks obtained by measuring the cell filled with superfluid $^4$He.

The whole cell is mounted on the center of a flange connected to the mixing chamber of the RIKEN rotating dilution refrigerator\cite{takahashi2006new}, the same apparatus as TO measurements by Choi \textit{et al}.\cite{Choi_and_Takahashi_etal_Science_2010,Choi_and_Takahashi_etal_PRL_2012,Choi_etal_PRB_2012}.
The refrigerator is rotated with angular velocity $-4 \le \Omega \le 4$ rad/s.
Since the measured moduli had no dependence on the direction of rotation, we show results obtained at $0 \le \Omega \le$ 4 rad/s.
In this work, we employed two polycrystalline solid $^4$He samples of pressure $P = 3.6$ and 5.4 MPa grown by the blocked capillary method from commercial $^4$He gas with $^3$He concentration $x_3$ = 0.3 ppm.

\section{Results\label{Results}}
\subsection{Effect of rotation on shear modulus}
\begin{figure} 
\includegraphics[width=0.45\textwidth]{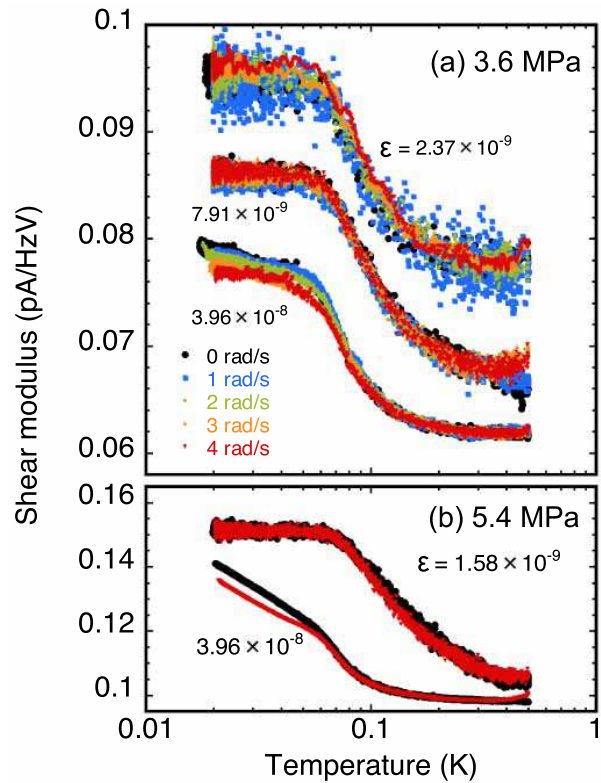}
\caption{(Color online)  
Shear modulus of two solid samples as a function of temperature. Data are shifted so as to coincide at 300 mK for each strain.
Scales of shear modulus for two samples are different.
(a) $P = 3.6$ MPa. Data are shown for three different strains and $\Omega = 0, 1, 2, 3, 4$ rad/s. 
(b) $P = 5.4$ MPa, for two strains and $\Omega = 0$ and 4 rad/s.
\label{fig:shear_all}}
\end{figure}

\begin{figure}
\includegraphics[height=0.5\textwidth]{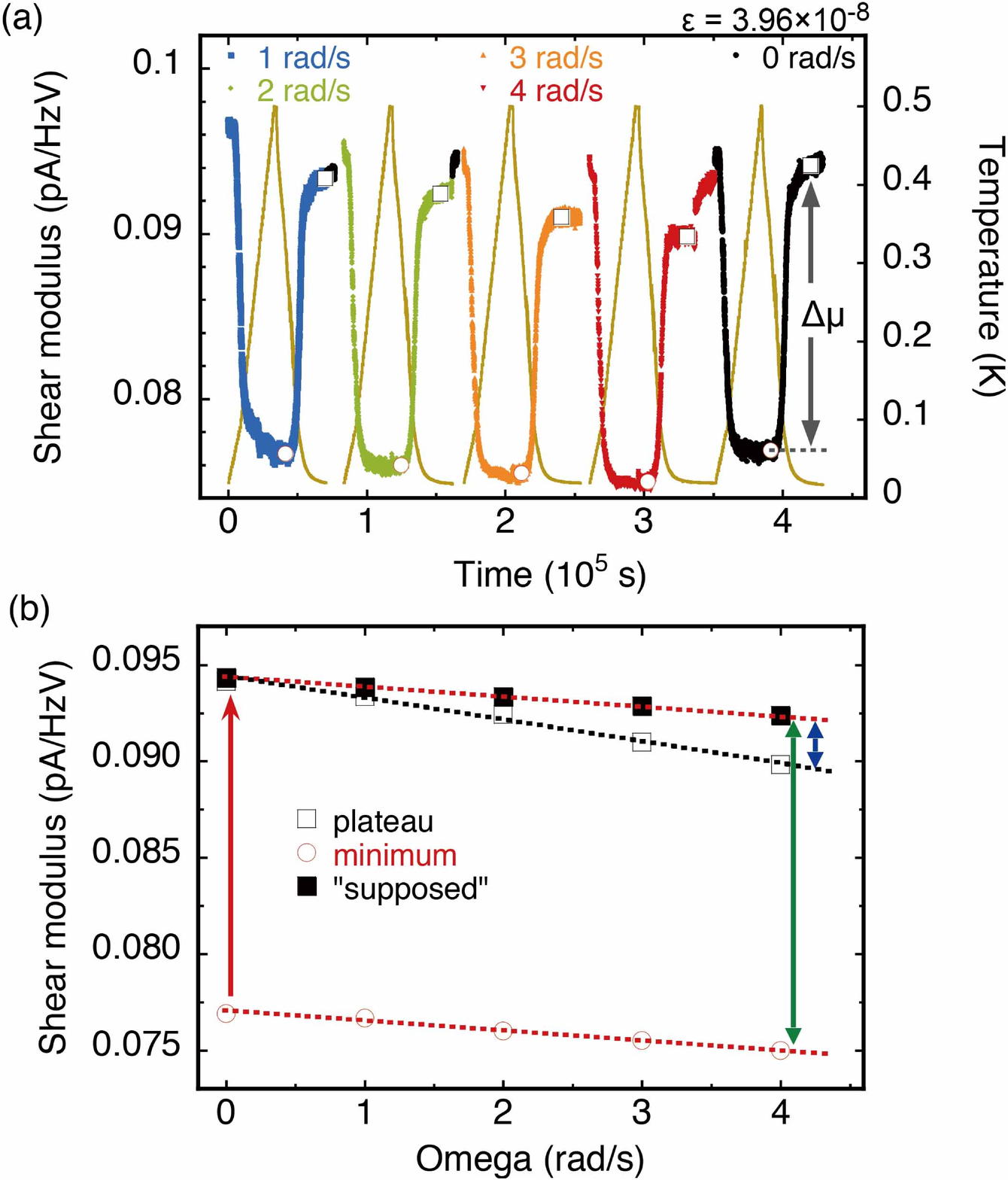}
\caption{(Color online) (a) A sequence of measurement of $\Omega$ dependent shear modulus $\mu$ at $\epsilon$ = 3.96 $\times 10^{-8}$. Brown lines show temperature, while other colors show $\mu$ at different $\Omega$'s. Black points indicate data at $\Omega = 0$ rad/s between 1 and 2 rad/s, 2 and 3 rad/s, after 4 rad/s measurements. Shear modulus has a minimum around 300 mK and has a plateau as the lowest temperature approaches. (b) $\Omega$ dependence of the "minimum" and the "plateau" values of $\mu$ ($\mu_{\rm ave}(300\ {\rm mK})$ and $\mu_{\rm ave}(20\ {\rm mK})$). The lowermost (red) and middle (black) dashed lines are results of the linear least square fitting. The uppermost (red) line is a parallel shift of the minimum line (indicated by red arrows), thus shows a "supposed" change in $\mu$ at the lowest T if there were no suppression by rotation.
Suppression of $\mu$ by rotation is described by blue arrow (see text).\label{fig:shear_36bar_time_current}}
\end{figure}

Figure \ref{fig:shear_all} shows the temperature dependence of shear modulus $\mu$ measured at three different strains ($\epsilon$) for two solid $^4$He samples of 3.6 and 5.4 MPa.
In Fig. \ref{fig:shear_all}, data for different $\Omega$'s are shifted to coincide at 300 mK on the assumption that all dislocation is free from pinning by $^3$He impurities so that they have essentially the same minimum modulus value at this temperature 
and the whole datasets with the same strain results are systematically shifted for clarity.

The data were taken as a time series (partly shown in Fig. \ref{fig:shear_36bar_time_current}) in the cooling run from 0.5 K to the lowest $T$.
In both samples, changes in $\mu$ depend on $T$ and $\epsilon$.
In the case of the 3.6 MPa sample shown in Fig. \ref{fig:shear_all} (a), at $\epsilon \le 7.91\times10^{-9}$, the saturation of change in $\mu$ completes at 40 mK. The modulus increases from 300 to 20 mK by about 22 $\%$.
On the other hand, at $\epsilon=3.96\times 10^{-8}$, $\mu$ does not saturate at 40 mK.
This is also observed in the sample of 5.4 MPa shown in Fig. \ref{fig:shear_all}(b), in which the magnitude of the change by temperature is about 40 $\%$ at two different $\epsilon$'s and at 0 rad/s.
This $\epsilon$ dependence is essentially similar to the previous shear modulus measurements by other groups\cite{Day_and_Beamish_Nature_2007,Day_etal_PRB_2009,Choi_and_Takahashi_etal_Science_2010,DYKim_etal_PRB_2011} using a pair of flat PZT. 
The trend that the saturation temperature shifts to lower $T$ at large $\epsilon$ indicates that $\epsilon=3.96\times 10^{-8}$ exceeds, or is in the vicinity of the critical strain $\epsilon_{\mathrm c}$, above which $\mu$ is suppressed. 

We examined the effect of rotation for the two samples. 
In the 3.6 MPa sample at $\epsilon = 2.37 \times 10^{-9}$, 
no characteristic dependence of $\mu$ on $\Omega$ is observed, although data are scattered.
At $\epsilon = 7.91 \times 10^{-9}$, $\mu$ does not change by rotation, except above 0.36 K, where the slope of $\mu$ changes as $\Omega$ increases to 2 rad/s.
Also in the 5.4 MPa sample, at $\epsilon = 1.58 \times 10^{-9}$, $\mu$ is almost identical for 0 and 4 rad/s, except for a small difference between 0.1 and 0.2 K.
Although the change is too small to identify as a rotation effect because of poor signal-to-noise ratio of the data, we will mention in Sec. \ref{Interpretation} that a similar decrement at intermediate temperatures (0.1 - 0.2 K) has been observed in the floating-core TO\cite{TsuikiPreparation}. 
At $\epsilon = 3.96\times 10^{-8}$, $\mu$ at two pressures tend to have smaller value below 0.1 K as $\Omega$ increases from 0 to 4 rad/s. 
This tendency is more prominently seen in the 5.4 MPa sample.

The $\Omega$ - dependent shear modulus at $\epsilon = 3.96\times 10^{-8}$ is more clearly confirmed by looking at the time sequence of the data acquisition. The time sequence of the 3.6 MPa data is shown in Fig. \ref{fig:shear_36bar_time_current} (a). 
We set $\Omega$ to a constant value (1, 2, 3 and 4 rad/s) before either cooling or warming scan.
After each scan at a constant $\Omega$, we stopped rotation ($\Omega$ = 0 rad/s) and then immediately increased $\Omega$ to a new value.
The open circles indicate the high - $T$ shear modulus which is obtained by averaging the data around 300 mK, and the open squares indicate the shear modulus averaged at the lowest $T$.
With these averaged data $\mu_{\mathrm {ave}}$, we define $\Delta \mu \equiv \mu_{\mathrm {ave}}(20\ {\mathrm {mK}}) - \mu_{\mathrm {ave}}(300\ {\mathrm {mK}})$, the magnitude of total change in $\mu$ from 300 to 20 mK, which is indicated by an arrow in Fig. \ref{fig:shear_36bar_time_current} (a) for the $\Omega$ = 0 rad/s data. $\Delta \mu$ decreases progressively as $\Omega$ increases.

There is another important feature in Fig. \ref{fig:shear_36bar_time_current}(a).
At the lowest $T$, $\mu$ once has a plateau value indicated by open square, then increases abruptly.
One can clearly see that this abrupt change occurs while keeping $\Omega = 4$ rad/s (shown in red data).
We will discuss this behavior in Sec. \ref{Interpretation}.

The $\Omega$ dependence of $\mu$ is obtained by plotting $\mu_{\mathrm {ave}}(300\ {\mathrm {mK}})$ and $\mu_{\mathrm {ave}}(20\ {\mathrm {mK}})$ (open circles and squares in Fig. \ref{fig:shear_36bar_time_current}(a)) as a function of $\Omega$. 
It is shown in Fig. \ref{fig:shear_36bar_time_current}(b). 
Shear moduli $\mu_{\mathrm {ave}}(300\ {\mathrm {mK}})$ and $\mu_{\mathrm {ave}}(20\ {\mathrm {mK}})$ are approximately linear in $\Omega$, having different gradients. 
From these averaged data we evaluate the rotation effect on shear modulus. 
We tentatively assume that the $\Omega$ dependence of $\mu$ at high $T$ is caused by the drift of measurement by unknown origin. 
We perform a least square linear fitting to this high - $T$ modulus, then utilize the fitting as a "supposed" shear modulus 
$\mu_{\mathrm {supposed}}(20\ {\mathrm {mK}})$ (shear modulus if there were no rotation effect) for the lowest $T$. 
These are shown in Fig. \ref{fig:shear_36bar_time_current}(b). 
The rotation effect is evaluated by a ratio $\delta(\Omega)$ defined as 

\begin{equation}
\delta(\Omega) \equiv \frac{\mu_{\mathrm {supposed}}(20\ {\mathrm {mK}}) - \mu_{\mathrm {ave}}(20\ {\mathrm {mK}})}{\mu_{\mathrm {supposed}}(20\ {\mathrm {mK}}) - 
\mu_{\mathrm {ave}}(300\ {\mathrm {mK}})}, 
\end{equation}
which is the ratio of the length of green arrow to blue one in Fig. \ref{fig:shear_36bar_time_current}(b). 
At $\Omega = 4$ rad/s, $\delta(4\ {\mathrm{rad/s}})$ is 0.147; i.e. the shear modulus is suppressed to 85 $\%$ of the original modulus without rotation. 
For the 5.4 MPa solid, $\delta(4\ {\mathrm{rad/s}})$ is 0.123, slightly smaller than that of 3.6 MPa case.
As to the definition of $\delta(\Omega)$, a different consideration is given in Appendix. 

\begin{figure}
\includegraphics[width=0.4\textwidth]{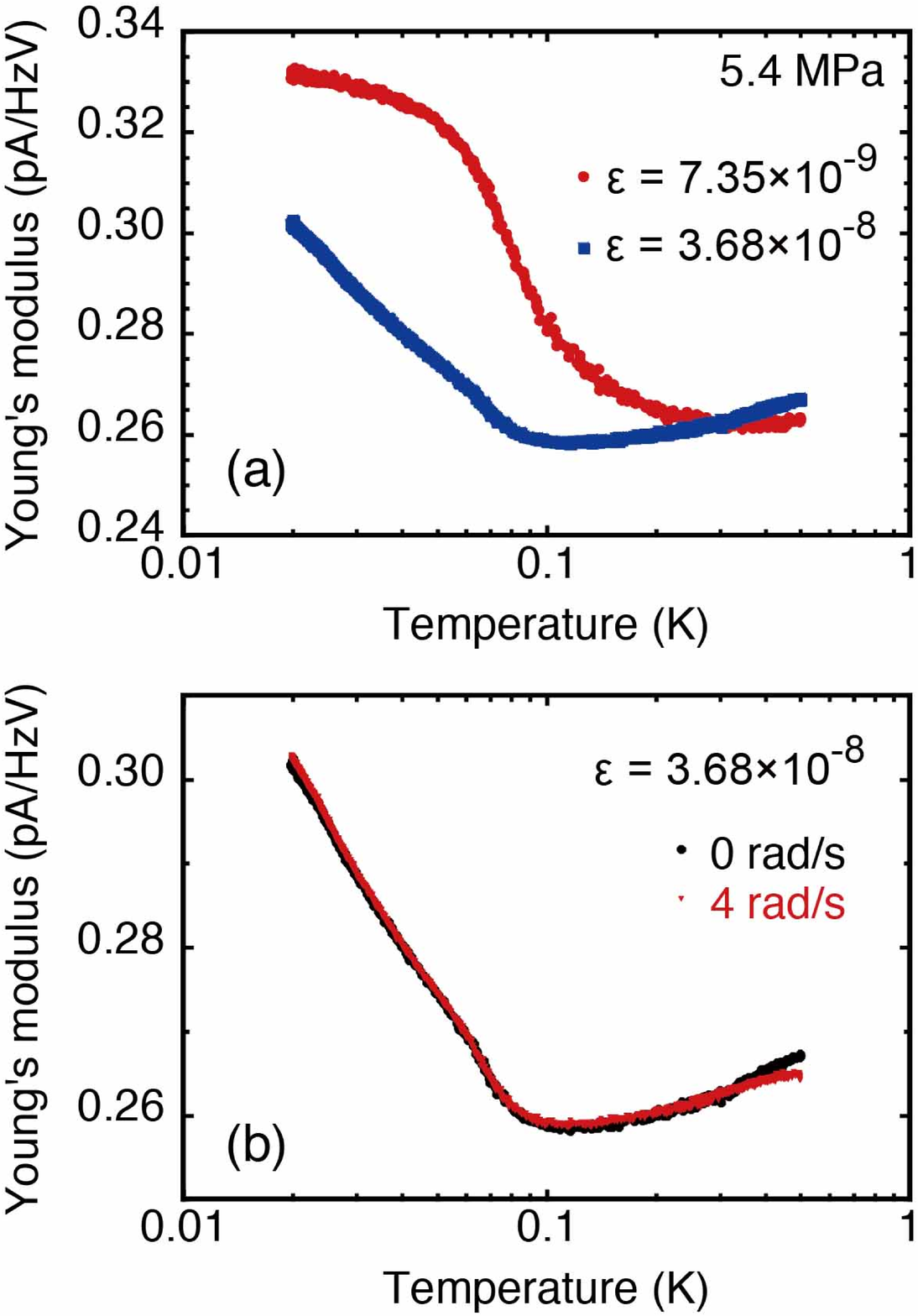}
\caption{(Color online) Young's modulus $E$ measured in cooling scans. (a) $E$ measured at $\epsilon = 7.35 \times 10^{-9}$ and $3.68 \times 10^{-8}$ without rotation. 
(b) $E$ measured at $\epsilon = 3.68 \times 10^{-8}$ with and without rotation. 
Data at $\Omega = 4$ rad/s are shifted to coincide at 300 mK.
\label{fig:comp_tscan_norm}}
\end{figure}

\subsection{Effect of rotation on Young's modulus}
Figure \ref{fig:comp_tscan_norm} (a) shows Young's modulus $E$ measured by the compressive PZT for 5.4 MPa sample at $\epsilon = 7.35 \times 10^{-9}$ and $3.68 \times 10^{-8}$ without rotation.
$E(T)$ gradually decreases as $T$ decreases, then shows a minimum, followed by a large increase.
At $\epsilon = 3.68 \times 10^{-8}$, the overall behavior shifts to lower temperatures, and $E$ does not saturate.
The change in $E$ from 200 to 20 mK is about 60 $\%$ of that of the lower strain case. 
By analogy with shear modulus, a critical strain for Young's modulus presumably exists between the two strains.
In Fig. \ref{fig:comp_tscan_norm}(b), we compare $E(T)$ data above the critical strain with and without rotation.
Unlike shear modulus, no significant difference was found in $E(T)$, except for a small downward shift at $\Omega = 4$ rad/s above 0.3 K.

\section{Interpretation for the rotation effect\label{Interpretation}}
In this Section, we discuss the possible mechanism for the rotation effect observed in shear modulus when strain over $\epsilon_{\mathrm c}$ is applied.
We firstly show that the observed rotation effect can occur when the probability of attaching $^3$He impurities to dislocations, $G$, decreases by three orders of magnitude. 
Next, we propose that such a change in $G$ can be realized by the effect of the Coriolis force exerted to $^3$He impuritons.

\subsection{Pinning of $^3$He to Dislocation Network}
Rotation produces centrifugal and Coriolis forces on the solid $^4$He samples, and these forces can influence both the structure of dislocation network and the motion of $^3$He impurities.
Iwasa points out that dislocation segments in the network can be stretched by the stress induced by rotation\cite{Iwasa_pc}. 
However, the magnitude of stretching by centrifugal force is estimated to be $10^{-12}$ of the segment length $L$, when $L$ is set to 1 $\mu$m. 
This stretching is too small to explain the 15 $\%$ reduction in $\mu$ by rotation (the effect of the Coriolis force is even smaller).

Next, we discuss the effect of rotation on the pinning of $^3$He to dislocations. 
In the dislocation pinning model\cite{Iwasa_JLTP_2013}, shear modulus $\mu$ is given by
\begin{eqnarray}
\mu &=& \frac{\mu_{\rm el}}{1 + \frac{(1-\nu)\Lambda}{2 \pi} L^2} \nonumber \\
&=& \mu_{\rm el} \kappa ^{-1} \quad \left( \kappa \equiv 1 + \frac{(1-\nu)\Lambda}{2 \pi} L^2 \right) \label{eq:mu_iwasa}
\end{eqnarray}
where $\mu_{\rm el}$ is the shear modulus when there is no pinning effect by impurity $^3$He, $\nu$ Poisson's ratio, $\Lambda$ the dislocation density, 
and $L$ the average length of dislocation between pinning points.
Hereafter, we adopt averages for all the lengths concerning dislocations for simplicity, but there are broad distributions in reality. 
We mention the effect of distributions later.

Equation (\ref{eq:mu_iwasa}) clearly indicates that the shear modulus decreases as $L$ increases.
$L$ is given by
\begin{eqnarray}
L=\frac{L_{\mathrm{NA}} L_{\mathrm{iA}}}{L_{\mathrm{NA}} + L_{\mathrm{iA}}} \label{eq:l}
\end{eqnarray}
where $L_{\mathrm{NA}}$ and $L_{\mathrm{iA}}$ are the \textit{average} of length of dislocation segment between network nodes, $L_{\mathrm{N}}$, and the \textit{average} of the dislocation length between the pinned points by $^3$He impurities, $L_{\mathrm{i}}$, respectively.
Since the dislocation nodes are strongly fixed, the distribution of $L_{\mathrm{N}}$ is determined when the solid sample is formed.
$L_{\mathrm{i}}$ is determined as $L_{\mathrm{N}}$ divided by the number of pinned $^3$He on a dislocation, $n_{\mathrm i}$, which is evaluated by considering pinning process.

When $^3$He impurities approach a dislocation segment, they pin the segment with pinning rate $R_1$ given by
\begin{eqnarray}
R_1 = x_3 L_{\mathrm{NA}} G \label{r_1}
\end{eqnarray}
where $x_3$ and $G$ are $^3$He concentration and the probability of pinning of $^3$He on the dislocation, respectively.
Pinned $^3$He is detached from the dislocation with a rate of unpinning rate $R_2$.
$R_2$ is obtained by assuming an Arrhenius type rate equation with thermal activation energy $E_{\mathrm A}$ by the binding energy of a $^3$He to dislocation as follows.
\begin{eqnarray}
R_2 = \frac{n_{\mathrm i}}{\tau_0}e^{-E_{\mathrm A}/T} \label{r_2}
\end{eqnarray}
where 
$\tau_0$ is the relaxation time ($1/\tau_0$ is the attempt frequency of the thermal activation of dislocation).
The equilibrium number of attached $^3$He on a dislocation $n_{\mathrm {i0}}$ is evaluated by balance of $R_1$ and $R_2$ so that we obtain $n_{\mathrm {i0}}$ given by
\begin{eqnarray}
n_{\mathrm i0} &=& x_3 L_{\mathrm {NA}} G \tau_0 e^{E_{\mathrm A}/T}.
\label{n_i0}
\end{eqnarray}
Eventually, $L_{\mathrm {iA}}$ is obtained:
\begin{eqnarray}
L_{\mathrm{iA}} = \frac{L_{\mathrm{NA}}}{n_{\mathrm i0}}=\frac{\exp{(-E_{\mathrm A}/T)}}{x_3 G\tau_0} \label{eq:L_i}.
\label{L_iA}
\end{eqnarray}

Taking Eqs. (\ref{eq:mu_iwasa}), (\ref{eq:l}) and (\ref{eq:L_i}) into account, we find that the quantity $\kappa$ depends on the pinning probability $G$,
\begin{eqnarray}
\kappa &=& 1 + \frac{(1-\nu)\Lambda}{2 \pi} \left(  \frac{L_{\mathrm{NA}} L_{\mathrm{iA}}}{L_{\mathrm{NA}} + L_{\mathrm{iA}}}\right) ^2 \label{eq:lambda} \\
&=& 1 + \frac{(1-\nu)\Lambda}{2 \pi} \left(  \frac{L_{\mathrm{NA}} \frac{\exp{(-E_{\mathrm A}/T)}}{x_3 G\tau_0}}{L_{\mathrm{NA}} + \frac{\exp{(-E_{\mathrm A}/T)}}{x_3 G\tau_0}}\right) ^2
\label{kappa}
\end{eqnarray}

We evaluate the magnitude of change in $\mu$ by changing temperature, using $\delta_\mu$, the same as Eq. (\ref{eq:definition}).
Using Eq. (\ref{eq:mu_iwasa}), Eq. (\ref{eq:definition}) becomes
\begin{eqnarray}
\delta_{\mu} &=& \frac{\frac{\mu_{\mathrm {el}}}{\kappa (0.02\ {\rm K})} - \frac{\mu_{\mathrm {el}}}{\kappa (0.3\ {\mathrm {K}})}}{\frac{\mu_{\mathrm {el}}}{\kappa (0.3\ {\mathrm {K}})}} \\
&=& \frac{\kappa (0.3\ {\mathrm {K}})}{\kappa (0.02\ {\mathrm {K}})}-1 
\label{eq:deltamu_lambda}.
\end{eqnarray}

According to Iwasa\cite{Iwasa_JLTP_2013}, we choose $\Lambda = 1.32\times 10^{10}$ m$^{-2}$, $G = 1.0\times 10^{12}$ m$^{-1}$s$^{-1}$, $\tau_0 = 10\times 10^{-3}$ s and $E_{\mathrm A}=0.2$ K.
Substituting these values to Eqs. (\ref{eq:L_i}), (\ref{eq:lambda}) and (\ref{eq:deltamu_lambda}), we obtain the expecting value for 3.6 MPa sample at 4 rad/s without suppression by rotation $\delta_\mu = 23.18 \%$ (in order to calculate this expected value, here $\mu_{\rm ave}(20\ {\rm mK})$ in $\delta_\mu$ is replaced by $\mu_{\rm supposed}(20\ {\rm mK})$ of the previous section), 
when we assume $L_{\mathrm{NA}}\approx1.36\times 10^{-5}\ {\rm m}$ according to Iwasa\cite{Iwasa_JLTP_2013}.

We now consider how rotation affects the change in $\delta_\mu$.
As previously mentioned, $L_{\mathrm{NA}} (=1.36\times 10^{-5}$ m) will not change by rotation.
Using Eqs. (\ref{eq:L_i}), (\ref{eq:lambda}) and (\ref{eq:deltamu_lambda}), it is deduced that the experimental value $\delta_\mu = 19.76 \%$ suppressed by rotation $\Omega = 4$ rad/s is fitted if we assume $G \approx 1.25\times 10^{9}$ m$^{-1}$s$^{-1}$. 
That is, by rotation, $G$ has to decrease $10^{-3}$ times the value of $G$ without rotation.

\subsection{Effect of rotation on the motion of $^3$He impurities}
In the previous section, we have concluded that the probability of sticking $^3$He to dislocations $G$ should decrease three orders of magnitude to explain the observed rotation effect on shear modulus. 
We propose that such a huge decrease in $G$ can be realized by the Coriolis force exerted to $^3$He impuritons. 
The rotational motion of the impuritons by the Coriolis force makes $^3$He diffusion strongly anisotropic between the direction of rotation and other directions. 

There are three types of forces exerted to rotating solid helium with constant rotation speed: gravity, centrifugal and Coriolis forces. 
Gravity and centrifugal forces are exerted to both the $^4$He and $^3$He atoms, and the magnitudes are $mg$ and $mr\Omega^2$, where $r$ is the radial position of the atom. 
Here the mass $m$ is the \textit{bare} mass of $^4$He or $^3$He. 
The centrifugal force can induce a distribution of $^3$He impurities in the radial direction, but the magnitude of the force at $\Omega = 4$ rad/s is an order of magnitude smaller than the gravitational force.
Therefore, the $^3$He density distribution caused by centrifugal force is negligible in the range of $\Omega$ in the present work.

On the other hand, the Coriolis force is a \textit{velocity - dependent force} exerted on an object moving with velocity $v$ in the rotating frame. 
Therefore, the Coriolis force is exerted \textit{only on moving} $^3$He \textit{impurities}, not on $^4$He atoms. 
The Coriolis force is given by $2m^* \vec{\Omega} \times \vec{v}$. 
We emphasize that the mass in the formula of Coriolis force is an \textit{effective mass} of $^3$He impuriton, and the velocity is the impuriton group velocity $v_{\mathrm {g}}$, which is determined by the energy bandwidth $\Delta$. 
Since the effective mass is large ($10^4 m_3 < m^* < 10^7 m_3$), the magnitude of the Coriolis force $2m^* \Omega v$ exceed several orders of magnitude of the gravitational and centrifugal forces. 

The equation of motion is dominated by the Coriolis force:
\begin{equation}
m^*\mathbf{a}=-2 m^*\mathbf{\Omega}\times\mathbf{v}.
\end{equation}

Consequently, the impuriton motion becomes \textit{spiral}, i.e. when we take the cylindrical coordinate ($r, \theta, z$) in which $z$ is the rotation axis, the motion is circular in the $r-\theta$ plane, whereas the motion in the direction of rotation axis ($z$) is unaffected. 
The radius of circle $R_{\mathrm{C}}$ is given by $\left< v_{\mathrm {g}}\right>_r/2\Omega$, where $\left< v_{\mathrm {g}}\right>_r$ is the projection of the group velocity vector to the $r-\theta$ plane. 
The magnitude of group velocity of $^3$He impuriton in hcp $^4$He is approximately given by\cite{huan2016quantum} $\left< v_{\mathrm {g}}^2\right>=18J_{34}^2 a^2 /\hbar^2$. 
Here $J_{34}$ is the exchange energy between $^3$He and $^4$He atoms, and $a$ is the lattice constant (of the hcp basal plane). 

In order to evaluate the rotation effect, we need to adopt values of $J_{34}$ from past NMR measurements, 
but no data are available for $^3$He concentration less than 10 ppm and for pressure higher than 4 MPa. 
We estimate $J_{34}$ from NMR studies of Greenberg \textit{et al}.\cite{greenberg1971isotopic,greenberg1972nuclear}, in which $J_{34}$ was obtained experimentally for solid samples with $^3$He concentration 0.01 and 0.02, and at molar volume from 21 to 20 cc/mole, which corrensponds to the pressure range from 2.5 MPa (i.e. melting pressure) to 4 MPa. 
Greenberg \textit{et al}. found that $\log J_{34}$ is linearly proportional to the molar volume $V_{\mathrm m}$, i.e. $J_{34} \propto \exp(V_{\mathrm m})$. 
Since no substantial $^3$He concentration dependence was seen in $J_{34}$, it is reasonable to assume the same order of magnitude for $J_{34}$ in our much more dilute samples (Note that the impuriton picture becomes better as the $^3$He concentration decreases). 
$J_{34}$ is estimated for our two samples by assuming the same molar volume dependence. 
We estimate possible maximum and minimum of $J_{34}$, taking the scattering of the original data of Greenberg into account. 
The lattice constant $a$ is estimated from the data of a past x-ray measurement\cite{shah_thesis_simmons}. 
Since $a$ is proportional to the cube root of the molar volume, it is reasonable to estimate $a$ at 3.6 and 5.4 MPa using least square fitting and using the data of pressure – molar volume relation obtained by Edwards and Pandorf\cite{edwards1965heat}. 

The estimated impuriton group velocity $v_{\mathrm{g}}$, the Coriolis force $F_{\mathrm{Cor}}$ and the radius of the impuriton circular motion $R_{\mathrm{C}}$ at $\Omega = 4$ rad/s are summarized in Table \ref{table1} together with the physical parameters used in the estimations. 
Note that the Coriolis force is a few orders of magnitude larger than gravitational and centrifugal forces ($\sim 10^{-26 \sim -28}$ N). 
The consequent circular motion of impuritons has a radius $R_{\mathrm{C}}$ about 1 and 10 $\mu$m at $P=5.4$ and 3.6 MPa, respectively. 

Under rotation, the motion of $^3$He impuritons will be composed of a circular motion in $r - \theta$ plane and a linear motion in $z$ direction. 
If the radius of circular motion $R_c$ is much smaller than the dislocation segment length $L_N$, the probability of $^3$He encountering to the dislocation $G$ will be suppressed.
Consequently, the shear modulus will decrease at low temperatures. 

 \onecolumngrid
\begin{widetext}

\begin{table}[tbh]
\begin{tabular}{|c|c|c|c|c|c|c|c|c|c|c|c|} \hline
\multicolumn{1}{|c}{$P$} & \multicolumn{1}{|c}{$V_m$} & \multicolumn{1}{|c}{$a$} & \multicolumn{2}{|c}{$f_{34}$ (kHz)} & \multicolumn{2}{|c}{$J_{34}$ (Joule)} & \multicolumn{2}{|c}{$v_{\mathrm{g}}$ (m/s)} & \multicolumn{1}{|c}{$F_{\mathrm{Cor}}$(N)} & \multicolumn{2}{|c|}{$R_{\mathrm{C}}$(m)} \\ \hline
(MPa) & (cc/mole) & (nm) & min & max & min & max & min & max & \, & min & max \\ \hline
2.50 & 21.17 & 0.3677 & $86$ & $100$ & $5.7\times10^{-29}$ & $6.6\times10^{-29}$ 
& $8.4\times10^{-4}$ & $9.8\times10^{-4}$ & $4.87\times10^{-24}$ & $1.0\times10^{-4}$ & $1.2\times10^{-4}$ \\ \hline
3.60 & 20.27 & 0.3624 & 6.93 & $10$ & $4.6\times10^{-30}$ & $6.6\times10^{-30}$ & $6.7\times10^{-5}$ 
& $9.7\times10^{-5}$ & $4.94\times10^{-24}$ & $8.4\times10^{-6}$ & $1.2\times10^{-5}$ \\ \hline
5.40 & 19.44 & 0.3574 & 0.35 & 2.0 & $2.3\times10^{-31}$ & $1.3\times10^{-30}$ & $3.3\times10^{-6}$ 
& $1.9\times10^{-5}$ & $5.01\times10^{-24}$  & $4.2\times10^{-7}$ & $2.4\times10^{-6}$ \\ \hline
\end{tabular}
\caption{Parameters and estimates of some quantities for three pressures: Pressure $P$, molar volume\cite{edwards1965heat} $V_{\mathrm m}$, lattice constant\cite{shah_thesis_simmons} $a$. Exchange frequency $f_{34}$ and bandwidth $J_{34}$ are obtained from Greenberg \textit{et al}.\cite{greenberg1971isotopic,greenberg1972nuclear}. 
Group velocity $v_{\mathrm{g}}$, the Coriolis force $F_{\mathrm{Cor}}$, and radius of circular motion $R_{\mathrm{C}}$ are derived from $J_{34}$ and $a$.
$F_{\mathrm{Cor}}$ and $R_{\mathrm{C}}$ are the values at $\Omega = 4$ rad/s.
The minima and maxima for some quantities are anticipated from the $f_{34}$ data of Greenberg \textit{et al}., taking the scatters of the data into account. 
Note that $F_{\mathrm{Cor}}$ is same for the maximum and minimum of $v_{\mathrm{g}}$. \label{table1}}

\end{table}

\end{widetext}

\twocolumngrid
Two experimental situations should be further considered in this interpretation: 
(1) The suppression of $\mu$ is observed only when the applied shear strain exceeds the critical strain. 
(2) In any real crystals, there is a broad distribution in the segment length $L_{\mathrm N}$ and the length between impurity pinning points $L_{\mathrm {i}}$. 
We discuss the rotation effect involving these facts, based on the dislocation pinning model extended by Iwasa\cite{Iwasa_JLTP_2013}, Fefferman \textit{et al}.\cite{fefferman2014dislocation} and Kang. \textit{et al}.\cite{kang2015modified}.

\begin{figure}
\includegraphics[width=0.48\textwidth]{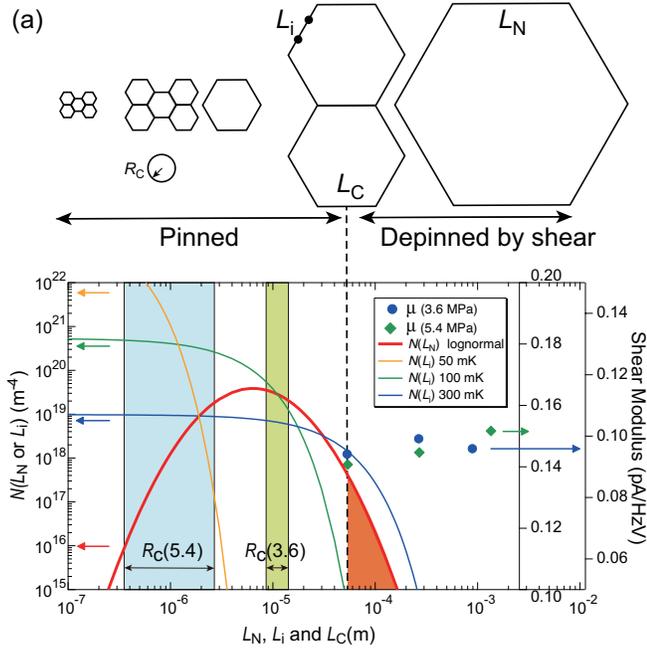}
\caption{(Color online) (a) Schematic illustration of distribution in $L_{\mathrm N}$, which corresponds to the abscissa of graph (b). $L_{\mathrm i}$ and $R_{\mathrm c}$ are also shown. (b) Number density of dislocation lengths $L_{\mathrm N}$ and $L_{\mathrm i}$, denoted as $N(L_{\mathrm N})$ and $N(L_{\mathrm i})$, as a function of $L_{\mathrm N}$ and $L_{\mathrm i}$ (for temperatures at 50, 100, and 300 mK), respectively. These are anticipated by Eqs. (\ref{N_L}) and (\ref{N_Li}). Measured shear modulus $\mu$ for the 3.6 and 5.4 MPa samples is shown as a function of $L_{\mathrm c}$, which is given by Eq. (\ref{L_c}). We focus on the location of $\mu$ at $L_{\mathrm c}=5\times 10^{-5}$m ($\epsilon = 3.96\times 10^{-8}$). The location is indicated by dashed line. The dislocation segment between nodes with lengths $L_{\mathrm N} > L_{\mathrm c}$ is depinned from $^3$He by applied shear stress. This is indicated as orange colored area of $N(L_{\mathrm N})$. $R_{\mathrm c}$ shown by colored (green and turquoise) area and double-headed arrows, which indicate uncertainty (see text), is much smaller than $L_{\mathrm c}$. Therefore, the pinning of $^3$He to dislocation segments at lengths $L_{\mathrm N} \le L_{\mathrm c}$ will be suppressed, resulting in the decrement in $\mu$. 
\label{LcandLn}}
\end{figure}

When a stress $\sigma$ is applied to crystal, dislocation strings feel a force. 
The force $f$ which needs to depin a $^3$He from dislocation has a critical value $f_{\mathrm c} = b \sigma L_{\mathrm c}/2$. 
Here $b$ is the Burgers vector and $L_{\mathrm c}$ is a critical segment length which separates pinned and depinned dislocations: 
i.e. dislocation segments shorter than $L_{\mathrm c}$ are pinned, while those longer than $L_{\mathrm c}$ are depinned.
$L_{\mathrm c}$ is therefore given by 
\begin{equation}
L_{\mathrm c} = 2f_{\mathrm c}/b\sigma,
\label{L_c}
\end{equation}
and is inversely proportional to applied stress $\sigma$.
For simplicity we assume that $L_{\mathrm c}$ is constant throughout one solid sample, although there might be inhomogeneity in its magnitude caused by distribution in the magnitude of activation energy $E_{\mathrm A}$.
$L_{\mathrm c}$ is obtained by using three shear stresses obtained in the measurement. 
As to the critical force, we adopt $f_{\mathrm c} = 6.8 \times 10^{-15}$ N, which was obtained by Fefferman \textit{et al}.\cite{fefferman2014dislocation}.
This value is close to other estimation $f_{\mathrm c} = 1 \times 10^{-14}$ N by Iwasa \textit{et al}.\cite{iwasa1979temperature}. 
We assume $b=a$, $a$ is shown in Table \ref{table1}.
In Fig. \ref{LcandLn} (b), we plot the measured shear modulus $\mu=I/fV(20 {\mathrm {mK}})$ as a function of $L_{\mathrm c}$. 

Once a solid sample is formed, $L_{\mathrm N}$ has a fixed distribution. 
It is appropriate to assume a lognormal distribution for $L_{\mathrm N}$\cite{iwasa1979temperature,kang2015modified}.
According to Kang \textit{et al}., the distribution of $L_{\mathrm N}$ is given by
\begin{eqnarray}
N(L_{\mathrm N})dL_{\mathrm N}=Z \exp\left[ - \frac{\left(\ln L_{\mathrm N} - \ln = L_{\mathrm N}\right)^2}{s^2}\right]dL_{\mathrm N},
\label{N_L}
\end{eqnarray} 
where $s$ is the width of the distribution. $Z$ and $= L_{\mathrm N}$ are $Z =\frac{\Lambda}{\sqrt{\pi}s{L_{\mathrm NA}}^2}\exp\left(\frac{s^2}{2}\right)$ and $= L_{\mathrm N}=L_{\mathrm NA}\exp \left(-\frac{3s^2}{4}\right)$, respectively.
On the other hand, an exponential distribution is assumed for $N(L_{\mathrm i})$,
\begin{equation}
N(L_{\mathrm i})dL_{\mathrm i}=\frac{\Lambda_{\mathrm d}}{{L_{\mathrm iA}}^2}\exp\left(-\frac{L_{\mathrm i}}{L_{\mathrm iA}}\right)dL_{\mathrm i},
\label{N_Li}
\end{equation} 
where we take thermal activation formula Eq. \ref{eq:L_i} with activation enerby $E_{\mathrm A} =  0.3$ K to obtain $N(L_{\mathrm i})$ at 50, 100 and 300 mK. 
Note that the choice of the value of $E_{\mathrm A}$, in which various authors claim different values (0.2 - 0.7 K), does not give any influence to our discussion. 
In both formula, $N(L_{\mathrm {N (or \: i)}})dL_{\mathrm {N (or \: i)}}$ is the number of dislocation segments having length between $L_{\mathrm {N (or \: i)}}$ and $L_{\mathrm {N (or \: i)}}+dL_{\mathrm {N (or \: i)}}$, 
and the integral $\int_0^\infty L_{\mathrm {N (or \: i)}}N\left(L_{\mathrm {N (or \: i)}}\right)dL_{\mathrm {N (or \: i)}}$ gives the total length of dislocations per unit volume $\Lambda$.
$N(L_{\mathrm N})$, and $N(L_{\mathrm i})$ for 50, 100 and 300 mK are shown in Fig. \ref{LcandLn} (b). 
In the calculation we take the average $L_{\mathrm NA}=1.36\times 10^{-5}$ m.

Let us first consider the virtual case that neither shear nor rotation is applied. 
In this case the pinning of $^3$He to dislocations are controlled only by temperature. 
In Fig. \ref{LcandLn}, as temperature decreases, the distribution $N(L_{\mathrm i})$ shifts to shorter $L_{\mathrm i}$ (left-handed) side, while $N(L_{\mathrm N})$ is unchanged.
This means that shorter dislocation segments are pinned by $^3$He at lower $T$. 
This is exactly what happens when solid $^4$He stiffens. 
 
Next, we think of the case that some shear strain is applied.
$N(L_{\mathrm N})$ (red curved line shown in Fig. \ref{LcandLn}(b)) has a peak at $L_{\mathrm N} \sim 7 \times 10^{-6}$ m, and decreases logarithmically as $L_{\mathrm N}$ increases.
When applied shear is weak and hence $L_{\mathrm c}$ is large, e.g. $L_{\mathrm c} > 1.5 \times 10^{-4}$ m, there are very few dislocation segments which are depinned by applied shear.
This is quite consistent with our observation that $\mu$ was not suppressed at shear strain $\epsilon \le 7.91\times 10^{-9}$, which corresponds to four data points of $I/fV$ at $L_{\mathrm c} \ge 3 \times 10^{-4}$ m of Fig. \ref{LcandLn}(b).

At the highest shear strain $\epsilon = 3.96\times 10^{-8}$, $L_{\mathrm c}$ is located at $5 \times 10^{-5}$ m, which is indicated by dashed line across Fig. \ref{LcandLn}(a) and (b).
In this case, the dislocations having lengths $L_{\mathrm N} > L_{\mathrm c}$, which are shown by orange colored region, are all depinned from $^3$He by applied shear, even if the distribution $N(L_{\mathrm i})$ went to short $L_{\mathrm i}$ side.
This depinning occuring in the long segments will result in the suppression of shear modulus, which is actually observed in measurement.

Let us consider the effect of rotation. 
We show in Fig. \ref{LcandLn}(b) the radius of circular motion $R_{\mathrm C}$ at $\Omega = 4$ rad/s.
The ambiguity in the estimation of $R_{\mathrm C}$ is expressed by colored area (green and blue for 3.6 and 5.4 MPa, respectively).
$R_{\mathrm C}$ is approximately $\frac{1}{5}$ and $\frac{1}{50}$ of $L_{\mathrm c}$ at $\epsilon = 3.96\times 10^{-8}$ (the dashed line) for pressure 3.6 and 5.4 MPa, respectively.
The probability of pinning of $^3$He to dislocation segments with lengths $L_{\mathrm N} \leq L_{\mathrm c}$ will be suppressed by the spiral motion of impuritons.
It is therefore expected that shear modulus is suppressed by rotation, even in the case that dislocation segments has a distribution in lengths.  

In this explanation, the necessary condition for the suppression of pinning probability $G$ at a certain temperature $T$ is that $R_{\mathrm C}$ is much smaller than dislocation segment length which mainly contribute to shear modulus at $T$. 
Once the distribution $N(L_{\mathrm N})$ is fixed in a particular solid sample, the magnitude relationship between $R_{\mathrm C}$ and $L_{\mathrm N}$ is also fixed. 
If $R_{\mathrm C} \ll L_{\mathrm N}$ holds for most of the dislocation segments (as in the case of Fig. \ref{LcandLn}(b)), the suppression of $\mu$ will occur irrespective of applied shear stress. Nevertheless, why is the rotation effect is seen only at $\epsilon \ge \epsilon_{\mathrm c}$?
We consider that the suppression of $\mu$ exists also in the data of smaller shear strains, $\epsilon < 10^{-8}$.
In the $\mu$ data of the 5.4 MPa sample, $\mu$ at $\Omega = 4$ rad/s is slightly smaller than $\mu$ at rest in the temperature range between 0.1 and 0.2 K.
It is currently difficult to conclude whether this decrease in $\mu$ is the true rotation effect or not, because the change is close to the experimental resolution of shear modulus.
However, we have observed similar changes in a floating-core TO mentioned in Sec. \ref{Backgrounds}, in which the TO resonant frequency decreases accompanied with additional dissipation in the temperature range between 50 and 300 mK, depending on the TO amplitude, which is roughly proportional to applied shear strain. 
The floating-core TO is so sensitive to the change in shear modulus of solid He contained between the inner core and outer shell that the rotation effect in shear modulus could be detected even if it were small. The TO data are now under analysis and will be published elsewhere. 

\begin{figure}
\includegraphics[width=0.3\textwidth]{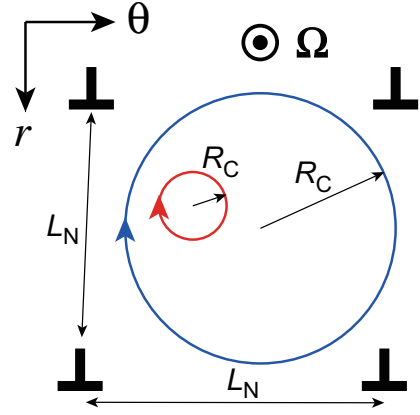}
\caption{
(Color online) Schematic view of dislocation network and circular impuriton motion viewed in the direction of rotation axis $z \parallel \Omega$. 
As discussed in Sec. \ref{Backgrounds}, edge dislocation lines parallel to $z$ mostly contribute to $\mu$ when shear is applied to $\theta$ direction. 
For simplicity, four dislocation lines are shown as $\perp$ keeping distance $L_{\mathrm N}$. 
When $R_{\mathrm c} \ll L_{\mathrm N}$, impuritons do not encounter the dislocations, 
and sticking such impuritons to dislocation lines in $r$ and $\theta$ 
directions (not shown) do not contribute significantly to the change 
in $c_{44}$. 
Impuritons rotating with $R_{\mathrm c} \sim L_{\mathrm N}$ will be captured by the $z$ - dislocations, so the rotation effect will disappear as molar volume of solid $^4$He increases.
\label{Dislocation_Coriolis}}
\end{figure}

Our interpretation for the rotation effect can be reinforced by considering directional anisotropy in dislocation segments that contribute to shear modulus.
Figure \ref{Dislocation_Coriolis} shows a cartoon of a part of dislocation network and circular impuriton motion, which are viewed from top of the PZT pair, i.e. in the direction of rotation axis $z \parallel \Omega$. 
As explained in Sec. \ref{Backgrounds}, edge dislocation lines parallel to $z$ mostly contribute to $\mu$ when shear is applied to the $\theta$ direction. 
For simplicity, four such dislocation lines are shown as $\perp$ separated by distance $L_{\mathrm N}$: i.e. four lines are connected by other dislocation segments running in the directions of $r$ and $\theta$, which are not shown in the cartoon.
It is easily seen that, when $R_{\mathrm c} \ll L_{\mathrm N}$, impuritons do not encounter the dislocations.  
On the other hand, impuritons moving (spirally) in $z$ direction can still be trapped to dislocation lines lying in $r - \theta$ plane. 
However, pinnings of dislocation lines in $r - \theta$ plane do not contribute significantly to the change in $c_{44}$ measured by PZT driven in $\theta$ direction.
Therefore, the $z$-directed impuriton motion will not influence the rotation effect.
 
As easily understood in Fig. \ref{Dislocation_Coriolis}, impuritons rotating with $R_{\mathrm c}$ close to $L_{\mathrm N}$ will be eventually captured by dislocation lines in $z$ direction.
The rotation effect will therefore decline as molar volume (pressure) of solid $^4$He increases (decreases), and will finally disappear as the pressure approaches melting point, where $R_{\mathrm c}$ exceeds $1 \times 10^{-4}$m, within which the lengths $L_{\mathrm N}$ of most of the dislocation segments lie.
Such a molar volume dependence may also explain the abrupt changes in $\mu$ we have observed near the end of each cooling scan of the 3.6 MPa sample .

\section{Discussion\label{Discussion}}
Our observation of the suppression of shear modulus by steady rotation can be consistently related to the results of TO experiments under rotation. 
The interpretation based on circular motion of $^3$He impuritons is also consistent with a theory of defecton quasiparticles in rotating quantum crystal.
Here we compare our results with these previous experiments and theories, and comment on some unsolved problems.

\subsection{Absence of rotation effect on Young's modulus}
We have not observed significant change in Young's modulus under rotation. 
Within the consideration of elasticity of continuous media, the absence of the rotation effect in Young's modulus is puzzling, since shear and Young's moduli are mutually related by Poisson's ratio. 
The absence of rotation effect might arise from inhomogeneity in displacements occuring in compression (expansion) experiment.
Our solid helium sample in measurement is long in azimuthal ($\theta$) direction while it is thin in between two PZTs. 
Moreover, the solid near the edge of PZT pair cannot freely expand because the solid exists also outside the sample between the PZTs, and because top and bottom of the solid sample is enclosed by BeCu wall. 
In this situation, compression of solid in $r$ direction will not produce much expansion in both $\theta$ and $z$ directions.
As a result the response detacted by one of the compression PZTs may only be determined by pure compression or expansion in $r$ direction but will not contain the effect of shear motion in $\theta$ direction. 

Simulation such as FEM will help to elucidate the validity of this speculation.

\subsection{Comparison with rotating torsional oscillator experiments}
The suppression of shear modulus by rotation observed in the present work has a resemblance to the results of rotating TO experiments performed by various groups\cite{yagi2011probable,choi2017superfluid,TsuikiPreparation}.
We propose that the most of the TO behaviors induced by steady rotation are essentially caused by suppression of shear modulus.

Yagi \textit{et al}. employed a TO containing a disk-shape solid $^4$He (formed from a commercial $^4$He gas, solid pressure not exactly known) and rotated it up to 1.26 rad/s (0.2 rps) with TO driving velocity up to 400 $\mu$m/s\cite{yagi2011probable}.
The shift of resonant frequency, which is now attributed to stiffening of solid $^4$He inside the TO bob, decreased as the rotation was applied below 70 mK, although the authors were not aware of the decrement.
The decrement became more prominent as $\Omega$ increased, and as the oscillation velocity of TO, $v_{\mathrm {rim}}$ (velocity of rim of the solid sample) increased:
at $v_{\mathrm {rim}} = 400 \mu$m/s, which was their highest velocity, the frequency shift caused by stiffening of $^4$He is suppressed by approximately 10 \% of from the original magnitude under no rotation.

Jaewon Choi \textit{et al}. studied the effect of rotation for a TO which has two annular tubes for solid $^4$He samples, and which has two resonant frequencies\cite{choi2017superfluid}.
When rotation was applied, decrement of period shift was observed in lower resonant mode below 80 mK. 
The decrement was roughly proportional to $\Omega$ and was about 15 \% when rotation $\Omega=4$ rad/s was applied. 
The high resonant mode had no significant rotation effect.

The rotation effects observed by these two TOs are attributed to the change in shear modulus.
As mentioned in Sec. \ref{Backgrounds}, the change in TO period are attributed to the change in shear modulus of solid $^4$He inside the TO bob.
Given that the low - $T$ period shift from the period at high $T$ (e.g. 500 mK) is entirely originated from stiffening of solid $^4$He, 
the period shift $\Delta P(T)$ can be considered to be proportional to the change in shear modulus $\Delta \mu(T)$.
The magnitude of the rotation - induced decrement of $\mu$ is close to our result.
This suggests that the distributions of dislocation network length $L_{\mathrm N}$ are identical in these two TOs and in our shear modulus apparatus.

In comparison of TO experiments with shear modulus measurements, the estimation of magnitude of shear strain or stress produced by oscillation of torsion bob could be a problem:
A large discrepancy has been pointed out in the nonlinear behaviors of $\Delta \mu(\epsilon)$ in direct shear measurements and $\Delta P(\epsilon)$ in TO studies.
This might be caused by inaccuracy in the estimation of $\epsilon$ produced by TO, which is described in Appendix.
In particular, the magnitude of the critical pinning length $L_{\mathrm c}$ is not clear in TO.

Another TO experiment by Fear \textit{et al}. found no significant effect in $\Delta P$ at 28 mK when cryostat was rotated up to 2 rad/s\cite{fear2016no}. 
This was probably due to the small magnitude of $\Delta P(T)$ in their TO design, which made the observation of even smaller rotation effect difficult.

Tsuiki \textit{et al}. have recently carried out a rotation experiment using a floating-core TO\cite{TsuikiPreparation}. 
This TO has a number of advantages for studying shear stiffening of solid $^4$He:
It is very sensitive to the change in shear modulus of solid He between inner metal core and outer shell, which can oscillate in phase and out of phase, and can produce extremely large frequency shift,
which is about 0.3 Hz at low resonant mode ($f \sim 850$ Hz) and about 50 Hz at higher modes ($f \sim 6300$ Hz). 
These features enables us to perform detailed studies of rotation effect on shear modulus of solid $^4$He with highest sensitivity we have ever had.
In this floating-core TO, Tsuiki \textit{et al}. have observed rotation - induced suppression of frequency shift at two resonant frequencies (about 850 and 6300 Hz), but the suppression most prominently occured at temperatures between 50 and 200 mK (depending on applied TO drive), and the suppression below 50 mK was not as large as the case of other TOs.
We suppose that this difference are attributed to the difference in the distribution of $L_{\mathrm N}$. 
In the floating-core TO $L_{\mathrm N}$ might have broader distribution than other TOs and our present apparatus, and the distribution might incline to long $L_{\mathrm N}$ side.
Such a distribution can shift the occurence of suppression of $^3$He pinning to high temperature side.

These TO results are consistent with each other and with our present work.
However, they do not reproduce the result of Choi \textit{et al}.\cite{Choi_and_Takahashi_etal_Science_2010,Choi_and_Takahashi_etal_PRL_2012,Choi_etal_PRB_2012}:
In Choi's experiment, when rotation $\Omega = 4$ rad/s was applied, the decrement of $\Delta P(T)$ reached 42 \% of $\Delta P(T)$ at $\Omega = 0$.
It is still difficult to explain such a large rotation effect by simply applying the suppression of shear modulus caused by circular motion of $^3$He impuritons. 
Similar large rotation effect was observed in a TO containing ring - shaped porous Vycor glass (8 mm ring radius, 2 mm thick in radial direction, and 12 mm in height)(Tsuiki \textit{et al.}, in preparation\cite{TsuikiPreparation}).
In this TO, narrow space of about $30 \mu$m thick was probably formed between the Vycor ring and the metal outer wall, due to the difference of thermal contraction in Vycor glass and BeCu metal,
and solid $^4$He confined in the narrow space produced large period shift (note that solid $^4$He in Vycor nanopores does not show stiffening because of absence of dislocations in solid $^4$He confined in nanopores). 
This large period shift showed a large decrement by rotation below 300 mK.
Surprisingly, the decrement at $\Omega = 4$ rad/s reached 60 \% of the original period shift. 

We speculate that these large rotation - induced decrements are originated from solid $^4$He in narrow spaces such as cracks or crevices in the torsion bob.
Such narrow spaces, for example, could be unintentionally formed in epoxy resin (Stycast etc.) which glues metal parts. 
It has been proposed in several experiments that stiffening of solid $^4$He existing inside such cracks gives nonnegligible contribution to the period shift of TO\cite{kim2014upper}.
In particular, if layers of thin solid $^4$He with thickness larger than $10 \mu$m (larger than $L_{\mathrm NA}$) are located closely to the place of torsion rod,
their stiffening by $^3$He - dislocation pinning mechanism produces a large period shift at low temperatures\cite{maris2012effect,kim2012absence,kim2014upper}.
If dislocations are formed more in narrow spaces than in larger (i.e. bulk) open spaces, enough $^3$He atoms necessary for completely pin the dislocations may not be provided from larger spaces when TO is rotated, because the diffusion of $^3$He is strongly disturbed by the rotation - induced Coriolis force. 
This scenario of nonequilibrium $^3$He spatial distribution may explain the large rotation effect of Choi \textit{et al}. and Tsuiki \textit{et al}., in which case was not taken for unintentional formation of narrow spaces.

\subsection{Theory of rotation effect in quantum solids}
As mentioned in Sec. \ref{Backgrounds}, Pushkarov theoretically studied the effects of rotation on quantum properties of solid $^4$He\cite{Pushkarov,pushkarov2001quasiparticle,pushkarov2012vacancy}. 
He proposed a Fokker - Planck type equation for defectons in solid, and derived the temperature dependence of diffusion coefficient tensor in the temperature regime that phonon scattering deterimines the defecton diffusion.
He obtained a surprising conclusion that diffusion coefficient $D$ in the directions perpendicular to the rotation axis change from $T^{-7}$ at rest and at small $\Omega$ to $T^9$ at $\Omega \sim 10^2$ rad/s, while $D$ in $z$ direction is unchanged at all. 
This change in the powerlaw exponent of $D_{r, \theta}$ makes defecton diffusion extremely anisotropic, and results in the same effect as we propose. 
In Pusukarov's theory the effect of rotation is introduced by a term proportional to $\Omega$ in collision integral, which is a consequence of Coriolis force acting on moving defectons.  
Since Pushkarov focused on zero - point vacancies (vacancion) in solid $^4$He, in which the effective mass is nearly the same order of magnitude as bare $^4$He mass,
the application of the theory to the case of $^3$He impuritons with large effective mass was not explicitly discussed.
In order to apply this theory to our experiment, the theory should be specified to the case of $^3$He impurities and be extended to the temperature regime where the effect of phonon scattering is negligible. 

We point out that future theories of quasiparticle dynamics taking into account the features of $^3$He impuritons (narrow bandwidth and heavy effective mass) and interaction between dislocations and impuritons will describe quantitatively the rotation effect on dislocation - $^3$He pinning.

\section{Conclusions\label{Conclusion}}
We have examined the effect of steady rotation on the elastic properties of solid $^4$He, using specially designed quarter circle PZTs and a rotating dilution refrigerator.
At low strains, no significant rotation effect was observed in shear modulus $\mu(T)$, except for a slight decrease in shear modulus around 200 mK in the sample of pressure $P = 5.4$ MPa. 
At strains presumably higher than or in the vicinity of the critical strain, above which $\mu$ decreases because the shear strain unbinds dislocations from $^3$He impurities, $\mu$ is suppressed by rotation below 80 mK when sample is rotated with angular velocity up to $\Omega = 4$ rad/s.
The decrement of $\mu$ at $\Omega = 4$ rad/s is about 14.7(12.3) $\%$ of the total change of $\mu(T)$ from 15 to 500 mK, for solid samples of pressure 3.6(5.4) MPa.
In order to explain the decrement, the probability of pinning of $^3$He on dislocation segment, $G$, must decrease about three orders of magnitude. 
Such a decrease in $G$ can be realized by the anisotropy of the motion (diffusion) of $^3$He impuritons by the Coriolis force.
We propose that the radius of the spiral motion of $^3$He impuritons, $R_{\mathrm c}$, can be much smaller than the length of dislocation segment $L_{\mathrm N}$. 
This will result in a decrease in pinning probability $G$ for dislocation lines aligned parallel to the rotation axis.

Our experiment and interpretation may solve the controversy in the results of several rotating torsional oscillator experiments.
Detailed analyses and discussion on several TO results including an analysis of the present shear modulus result by considering the distribution in $L_{\mathrm N}$ will be given in our future publication\cite{TsuikiPreparation}.

The microscopic mechanism of the rotation effect on $^3$He pinning is to be further investigated. 
Concerning to this problem, the dynamics of $^3$He impuritons is an interesting issue to pursuit.
Since the Coriolis force acting on $^3$He impuritons has a close analogy to the Lorentz force exerted to charged quantum particles\cite{dattoli2010note},
one may expect quantum phenomena as in rotating cold atoms\cite{cooper2008rapidly} and in electrons in magnetic field.  

\begin{acknowledgments}
We gratefully acknowledge fruitful discussions with Izumi Iwasa, Jaewon Choi and Eunseong Kim. 
D.T. acknowledges a financial support from Takahashi Industrial and Economic Research Foundation.
This work is supported by Grant-in-Aid for Scientific Research (S) (Grant No. 21224010) from JSPS and by RIKEN junior Research Associate Program.
\end{acknowledgments}

\appendix
\section{Evaluation of rotation effect}

{\bf (i)Another consideration of the decrement of shear modulus}

In the evaluation of the rotation - induced decrement of the shear modulus described in the main text, we assume that $\Omega$ dependence of $\mu$ at high $T$ appears due to the drift of measurement.
However, it is also possible to assume that the change in $\mu$ at high $T$ is a real rotation effect. 
In this case, it is not the absolute value of $\mu$ but the ratio of changes in $\mu$ from high to low $T$ that should be taken into consideration.
In this regard, we define a reduced change of shear modulus $\delta_{\mu}$ as follows:
\begin{eqnarray}
 \delta_{\mu} = \frac{\mu_{\mathrm {ave}}(20\ {\mathrm {mK}}) - \mu_{\mathrm {ave}}(300\ {\mathrm {mK}})}{\mu_{\mathrm {ave}}(300\ {\mathrm {mK}})} \label{eq:definition}
\end{eqnarray}
where $\mu_{\mathrm {ave}}(20\ {\mathrm {mK}})$ and $\mu_{\mathrm {ave}}(300\ {\mathrm {mK}})$ are defined in the main text and shown as open squares and circles  in Fig. 3 of the main text, respectively.

In Table \ref{table:modulus}, values of $\mu_{\mathrm {ave}}$ are summarized. 
Here we consider shear modulus at $\Omega = 0$ and 4 rad/s.
\begin{table}[!b]
	\caption{Shear modulus values of 3.6 MPa sample at low and high $T$, with and without rotation (unit : pA/HzV) {\label{table:modulus}}}
	\begin{tabular}{|c|c|c|} \hline
	 & 0 rad/s & 4 rad/s \\ \hline
	 $\mu_{\mathrm {ave}}(20\ {\mathrm {mK}})$ & 0.09415 & 0.08982 \\ \hline
	 $\mu_{\mathrm {ave}}(300\ {\mathrm {mK}})$ & 0.07689 & 0.07500 \\ \hline
	 \end{tabular}
\end{table}
The $\delta_{\mu}$ at 0 rad/s is evaluated to be 0.2245; i.e. 22.45 $\%$.
If there were no rotation effect in $\delta_{\mu}$, $\delta_{\mu}$ at 4 rad/s should be also 0.2245.
In reality, however, the results shown in Table \ref{table:modulus} show that $\delta_{\mu}$ at 4 rad/s is 0.1976 (19.76 $\%$), which is smaller than 0.2245.
Therefore, at $\Omega = 4$ rad/s $\delta _\mu$ decreases by 11.94 $\%$.
\\

{\bf (ii)Uncertainty in shear modulus}

\begin{figure}
\includegraphics[width=0.4\textwidth]{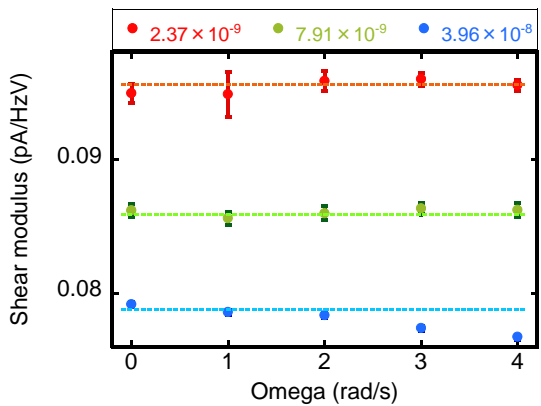}
\caption{(Color online) The average shear modulus $\mu$ with standard deviation as error bar for each $\Omega$ taken from Fig. \ref{fig:shear_all} in the main paper. Lines are guide to eye. \label{fig:omega_shear_noise}}
\end{figure}

Here we confirm the rotation effect on shear modulus by discussing uncertainty in shear modulus under rotation. 
We discuss the 3.6 MPa solid sample.
In Fig. \ref{fig:omega_shear_noise}, we show $\mu$ which is averaged at 20 mK for 30 minutes.
Standard deviations are indicated by error bars.
As shown in Fig. \ref{fig:omega_shear_noise}, at strains $\epsilon \le 7.91 \times 10^{-9}$, averaged shear modulus are on the horizontal line within error bar.
This indicates that $\mu$ at $\epsilon \le 7.91 \times 10^{-9}$ does not depend on $\Omega$.
On the other hand, 
$\mu$ at $\epsilon = 3.96 \times 10^{-8}$ decreases far over the error bars as $\Omega$ increases.
The decrease in $\mu$ at $\Omega = 4$ rad/s is an order of magnitude larger than the standard deviation.
Thus, the suppression of shear modulus is not experimental artifact but a significant rotation effect.

\section{Estimation of stress for annular solid $^4$He in a torsional oscillator}
In the previous TO experiment by Choi {\textit {\textit{et al}}}.\cite{Choi_and_Takahashi_etal_Science_2010}, suppression of TO period change by rotation was observed when they fixed AC oscillation speed at 6 $\mu$m/s 
(at the circumference of the annular sample).
By using this oscillation speed and other relevant values, the stress applied to solid $^4$He is calculated to be $\sim 1.0 \times 10^{-3}$ Pa.
The details of the calculation is shown below.

Considering the torque acting on solid $^4$He by the oscillation of the TO cell, we obtain the stress at sample as follows:
\begin{eqnarray}
	\sigma = \frac{\rho t \omega v}{2} \label{eq:to_sigma}
\end{eqnarray}
where $\rho$ is the density of solid $^4$He (for 4.0 MPa sample, $\rho \approx $160 kg/m$^3$), $t$ the thickness of the annular solid sample (0.4 mm), $\omega$ the resonant frequency ($2\pi \times 911$ rad/s), and $v$ the oscillation velocity of the sample.
Substituting $v = 6 \mu$m/s to Eq. (\ref{eq:to_sigma}), $\sigma$ is calculated to be about $1.0 \times 10^{-3}$ Pa, which is two orders magnitude smaller than $\sigma$ generated by the critical strain applied under present rotation experiment.


\begin{thebibliography}{56}%
\makeatletter
\providecommand \@ifxundefined [1]{%
 \@ifx{#1\undefined}
}%
\providecommand \@ifnum [1]{%
 \ifnum #1\expandafter \@firstoftwo
 \else \expandafter \@secondoftwo
 \fi
}%
\providecommand \@ifx [1]{%
 \ifx #1\expandafter \@firstoftwo
 \else \expandafter \@secondoftwo
 \fi
}%
\providecommand \natexlab [1]{#1}%
\providecommand \enquote  [1]{``#1''}%
\providecommand \bibnamefont  [1]{#1}%
\providecommand \bibfnamefont [1]{#1}%
\providecommand \citenamefont [1]{#1}%
\providecommand \href@noop [0]{\@secondoftwo}%
\providecommand \href [0]{\begingroup \@sanitize@url \@href}%
\providecommand \@href[1]{\@@startlink{#1}\@@href}%
\providecommand \@@href[1]{\endgroup#1\@@endlink}%
\providecommand \@sanitize@url [0]{\catcode `\\12\catcode `\$12\catcode
  `\&12\catcode `\#12\catcode `\^12\catcode `\_12\catcode `\%12\relax}%
\providecommand \@@startlink[1]{}%
\providecommand \@@endlink[0]{}%
\providecommand \url  [0]{\begingroup\@sanitize@url \@url }%
\providecommand \@url [1]{\endgroup\@href {#1}{\urlprefix }}%
\providecommand \urlprefix  [0]{URL }%
\providecommand \Eprint [0]{\href }%
\providecommand \doibase [0]{http://dx.doi.org/}%
\providecommand \selectlanguage [0]{\@gobble}%
\providecommand \bibinfo  [0]{\@secondoftwo}%
\providecommand \bibfield  [0]{\@secondoftwo}%
\providecommand \translation [1]{[#1]}%
\providecommand \BibitemOpen [0]{}%
\providecommand \bibitemStop [0]{}%
\providecommand \bibitemNoStop [0]{.\EOS\space}%
\providecommand \EOS [0]{\spacefactor3000\relax}%
\providecommand \BibitemShut  [1]{\csname bibitem#1\endcsname}%
\let\auto@bib@innerbib\@empty
\bibitem [{\citenamefont {Andreev}(1982)}]{Andreev}%
  \BibitemOpen
  \bibfield  {author} {\bibinfo {author} {\bibfnamefont {A.~F.}\ \bibnamefont
  {Andreev}},\ }in\ \href@noop {} {\emph {\bibinfo {booktitle} {Progress in Low
  Temperature Physics}}},\ Vol.\ \bibinfo {volume} {VIII},\ \bibinfo {editor}
  {edited by\ \bibinfo {editor} {\bibfnamefont {D.~F.}\ \bibnamefont
  {Brewer}}}\ (\bibinfo  {publisher} {North-Holland},\ \bibinfo {address}
  {Amsterdam},\ \bibinfo {year} {1982})\ pp.\ \bibinfo {pages}
  {67--131}\BibitemShut {NoStop}%
\bibitem [{\citenamefont {Pushkarov}(1991)}]{Pushkarov}%
  \BibitemOpen
  \bibfield  {author} {\bibinfo {author} {\bibfnamefont {D.~I.}\ \bibnamefont
  {Pushkarov}},\ }\href@noop {} {\emph {\bibinfo {title} {Quasiparticle Theory
  of Defects in Solids}}}\ (\bibinfo  {publisher} {World Scientific},\ \bibinfo
  {address} {Singapore},\ \bibinfo {year} {1991})\BibitemShut {NoStop}%
\bibitem [{\citenamefont {Varma}\ and\ \citenamefont
  {Werthamer}(1976)}]{Varma}%
  \BibitemOpen
  \bibfield  {author} {\bibinfo {author} {\bibfnamefont {C.~M.}\ \bibnamefont
  {Varma}}\ and\ \bibinfo {author} {\bibfnamefont {N.~R.}\ \bibnamefont
  {Werthamer}},\ }in\ \href@noop {} {\emph {\bibinfo {booktitle} {The Physics
  of Liquid and Solid Helium}}},\ \bibinfo {series and number} {Part I}\
  (\bibinfo  {publisher} {Wiley},\ \bibinfo {address} {New York},\ \bibinfo
  {year} {1976})\ pp.\ \bibinfo {pages} {503--570}\BibitemShut {NoStop}%
\bibitem [{\citenamefont {Guyer}\ and\ \citenamefont
  {Zane}(1970)}]{guyer1970mass}%
  \BibitemOpen
  \bibfield  {author} {\bibinfo {author} {\bibfnamefont {R.~A.}\ \bibnamefont
  {Guyer}}\ and\ \bibinfo {author} {\bibfnamefont {L.~I.}\ \bibnamefont
  {Zane}},\ }\href@noop {} {\bibfield  {journal} {\bibinfo  {journal} {Phys.
  Rev. Lett.}\ }\textbf {\bibinfo {volume} {24}},\ \bibinfo {pages} {660}
  (\bibinfo {year} {1970})}\BibitemShut {NoStop}%
\bibitem [{\citenamefont {Guyer}\ \emph {et~al.}(1971)\citenamefont {Guyer},
  \citenamefont {Richardson},\ and\ \citenamefont
  {Zane}}]{guyer1971excitations}%
  \BibitemOpen
  \bibfield  {author} {\bibinfo {author} {\bibfnamefont {R.~A.}\ \bibnamefont
  {Guyer}}, \bibinfo {author} {\bibfnamefont {R.~C.}\ \bibnamefont
  {Richardson}}, \ and\ \bibinfo {author} {\bibfnamefont {L.~I.}\ \bibnamefont
  {Zane}},\ }\href@noop {} {\bibfield  {journal} {\bibinfo  {journal} {Rev.
  Mod. Phys.}\ }\textbf {\bibinfo {volume} {43}},\ \bibinfo {pages} {532}
  (\bibinfo {year} {1971})}\BibitemShut {NoStop}%
\bibitem [{\citenamefont {Richards}\ \emph {et~al.}(1972)\citenamefont
  {Richards}, \citenamefont {Pope},\ and\ \citenamefont
  {Widom}}]{richards1972evidence}%
  \BibitemOpen
  \bibfield  {author} {\bibinfo {author} {\bibfnamefont {M.~G.}\ \bibnamefont
  {Richards}}, \bibinfo {author} {\bibfnamefont {J.}~\bibnamefont {Pope}}, \
  and\ \bibinfo {author} {\bibfnamefont {A.}~\bibnamefont {Widom}},\
  }\href@noop {} {\bibfield  {journal} {\bibinfo  {journal} {Phys. Rev. Lett.}\
  }\textbf {\bibinfo {volume} {29}},\ \bibinfo {pages} {708} (\bibinfo {year}
  {1972})}\BibitemShut {NoStop}%
\bibitem [{\citenamefont {Huang}\ \emph {et~al.}(1975)\citenamefont {Huang},
  \citenamefont {Goldberg},\ and\ \citenamefont {Guyer}}]{huang1975quantum}%
  \BibitemOpen
  \bibfield  {author} {\bibinfo {author} {\bibfnamefont {W.}~\bibnamefont
  {Huang}}, \bibinfo {author} {\bibfnamefont {H.~A.}\ \bibnamefont {Goldberg}},
  \ and\ \bibinfo {author} {\bibfnamefont {R.~A.}\ \bibnamefont {Guyer}},\
  }\href@noop {} {\bibfield  {journal} {\bibinfo  {journal} {Phys. Rev. B}\
  }\textbf {\bibinfo {volume} {11}},\ \bibinfo {pages} {3374} (\bibinfo {year}
  {1975})}\BibitemShut {NoStop}%
\bibitem [{\citenamefont {Esel'Son}\ \emph {et~al.}(1978)\citenamefont
  {Esel'Son}, \citenamefont {Mikheev}, \citenamefont {Grigor'ev},\ and\
  \citenamefont {Mikhin}}]{esel1978quantum}%
  \BibitemOpen
  \bibfield  {author} {\bibinfo {author} {\bibfnamefont {B.~N.}\ \bibnamefont
  {Esel'Son}}, \bibinfo {author} {\bibfnamefont {V.~A.}\ \bibnamefont
  {Mikheev}}, \bibinfo {author} {\bibfnamefont {V.~N.}\ \bibnamefont
  {Grigor'ev}}, \ and\ \bibinfo {author} {\bibfnamefont {N.~P.}\ \bibnamefont
  {Mikhin}},\ }\href@noop {} {\bibfield  {journal} {\bibinfo  {journal} {Sov.
  Phys. JETP}\ }\textbf {\bibinfo {volume} {74}},\ \bibinfo {pages} {2311}
  (\bibinfo {year} {1978})}\BibitemShut {NoStop}%
\bibitem [{\citenamefont {Andreev}\ and\ \citenamefont
  {Lifshits}(1969)}]{andreev1969quantum}%
  \BibitemOpen
  \bibfield  {author} {\bibinfo {author} {\bibfnamefont {A.}~\bibnamefont
  {Andreev}}\ and\ \bibinfo {author} {\bibfnamefont {I.}~\bibnamefont
  {Lifshits}},\ }\href@noop {} {\bibfield  {journal} {\bibinfo  {journal} {Sov.
  Phys. JETP}\ }\textbf {\bibinfo {volume} {56}},\ \bibinfo {pages} {2057}
  (\bibinfo {year} {1969})}\BibitemShut {NoStop}%
\bibitem [{\citenamefont {Chester}(1970)}]{chester1970speculations}%
  \BibitemOpen
  \bibfield  {author} {\bibinfo {author} {\bibfnamefont {G.~V.}\ \bibnamefont
  {Chester}},\ }\href@noop {} {\bibfield  {journal} {\bibinfo  {journal} {Phys.
  Rev. A}\ }\textbf {\bibinfo {volume} {2}},\ \bibinfo {pages} {256} (\bibinfo
  {year} {1970})}\BibitemShut {NoStop}%
\bibitem [{\citenamefont {Leggett}(1970)}]{leggett1970can}%
  \BibitemOpen
  \bibfield  {author} {\bibinfo {author} {\bibfnamefont {A.~J.}\ \bibnamefont
  {Leggett}},\ }\href@noop {} {\bibfield  {journal} {\bibinfo  {journal} {Phys.
  Rev. Lett.}\ }\textbf {\bibinfo {volume} {25}},\ \bibinfo {pages} {1543}
  (\bibinfo {year} {1970})}\BibitemShut {NoStop}%
\bibitem [{\citenamefont {Kim}\ and\ \citenamefont
  {Chan}(2004{\natexlab{a}})}]{kim2004probable}%
  \BibitemOpen
  \bibfield  {author} {\bibinfo {author} {\bibfnamefont {E.}~\bibnamefont
  {Kim}}\ and\ \bibinfo {author} {\bibfnamefont {M.~H.~W.}\ \bibnamefont
  {Chan}},\ }\href@noop {} {\bibfield  {journal} {\bibinfo  {journal} {Nature}\
  }\textbf {\bibinfo {volume} {427}},\ \bibinfo {pages} {225} (\bibinfo {year}
  {2004}{\natexlab{a}})}\BibitemShut {NoStop}%
\bibitem [{\citenamefont {Kim}\ and\ \citenamefont
  {Chan}(2004{\natexlab{b}})}]{kim2004observation}%
  \BibitemOpen
  \bibfield  {author} {\bibinfo {author} {\bibfnamefont {E.}~\bibnamefont
  {Kim}}\ and\ \bibinfo {author} {\bibfnamefont {M.~H.~W.}\ \bibnamefont
  {Chan}},\ }\href@noop {} {\bibfield  {journal} {\bibinfo  {journal}
  {Science}\ }\textbf {\bibinfo {volume} {305}},\ \bibinfo {pages} {1941}
  (\bibinfo {year} {2004}{\natexlab{b}})}\BibitemShut {NoStop}%
\bibitem [{\citenamefont {Ny{\'e}ki}\ \emph {et~al.}(2017)\citenamefont
  {Ny{\'e}ki}, \citenamefont {Phillis}, \citenamefont {Ho}, \citenamefont
  {Lee}, \citenamefont {Coleman}, \citenamefont {Parpia}, \citenamefont
  {Cowan},\ and\ \citenamefont {Saunders}}]{nyeki2017intertwined}%
  \BibitemOpen
  \bibfield  {author} {\bibinfo {author} {\bibfnamefont {J.}~\bibnamefont
  {Ny{\'e}ki}}, \bibinfo {author} {\bibfnamefont {A.}~\bibnamefont {Phillis}},
  \bibinfo {author} {\bibfnamefont {A.}~\bibnamefont {Ho}}, \bibinfo {author}
  {\bibfnamefont {D.}~\bibnamefont {Lee}}, \bibinfo {author} {\bibfnamefont
  {P.}~\bibnamefont {Coleman}}, \bibinfo {author} {\bibfnamefont
  {J.}~\bibnamefont {Parpia}}, \bibinfo {author} {\bibfnamefont
  {B.}~\bibnamefont {Cowan}}, \ and\ \bibinfo {author} {\bibfnamefont
  {J.}~\bibnamefont {Saunders}},\ }\href@noop {} {\bibfield  {journal}
  {\bibinfo  {journal} {Nature Physics}\ }\textbf {\bibinfo {volume} {13}},\
  \bibinfo {pages} {455} (\bibinfo {year} {2017})}\BibitemShut {NoStop}%
\bibitem [{\citenamefont {Day}\ and\ \citenamefont
  {Beamish}(2007)}]{Day_and_Beamish_Nature_2007}%
  \BibitemOpen
  \bibfield  {author} {\bibinfo {author} {\bibfnamefont {J.}~\bibnamefont
  {Day}}\ and\ \bibinfo {author} {\bibfnamefont {J.}~\bibnamefont {Beamish}},\
  }\href@noop {} {\bibfield  {journal} {\bibinfo  {journal} {Nature (London)}\
  }\textbf {\bibinfo {volume} {450}},\ \bibinfo {pages} {853} (\bibinfo {year}
  {2007})}\BibitemShut {NoStop}%
\bibitem [{\citenamefont {Day}\ \emph {et~al.}(2009)\citenamefont {Day},
  \citenamefont {Syshchenko},\ and\ \citenamefont
  {Beamish}}]{Day_etal_PRB_2009}%
  \BibitemOpen
  \bibfield  {author} {\bibinfo {author} {\bibfnamefont {J.}~\bibnamefont
  {Day}}, \bibinfo {author} {\bibfnamefont {O.}~\bibnamefont {Syshchenko}}, \
  and\ \bibinfo {author} {\bibfnamefont {J.}~\bibnamefont {Beamish}},\
  }\href@noop {} {\bibfield  {journal} {\bibinfo  {journal} {Phys. Rev. B}\
  }\textbf {\bibinfo {volume} {79}},\ \bibinfo {pages} {214524} (\bibinfo
  {year} {2009})}\BibitemShut {NoStop}%
\bibitem [{\citenamefont {Day}\ \emph {et~al.}(2010)\citenamefont {Day},
  \citenamefont {Syshchenko},\ and\ \citenamefont
  {Beamish}}]{Day_Syshchenko_and_Beamish_PRL_2010}%
  \BibitemOpen
  \bibfield  {author} {\bibinfo {author} {\bibfnamefont {J.}~\bibnamefont
  {Day}}, \bibinfo {author} {\bibfnamefont {O.}~\bibnamefont {Syshchenko}}, \
  and\ \bibinfo {author} {\bibfnamefont {J.}~\bibnamefont {Beamish}},\
  }\href@noop {} {\bibfield  {journal} {\bibinfo  {journal} {Phys. Rev. Lett.}\
  }\textbf {\bibinfo {volume} {104}},\ \bibinfo {pages} {075302} (\bibinfo
  {year} {2010})}\BibitemShut {NoStop}%
\bibitem [{\citenamefont {Beamish}\ \emph {et~al.}(2012)\citenamefont
  {Beamish}, \citenamefont {Fefferman}, \citenamefont {Haziot}, \citenamefont
  {Rojas},\ and\ \citenamefont {Balibar}}]{Beamish_etal_PRB_2012}%
  \BibitemOpen
  \bibfield  {author} {\bibinfo {author} {\bibfnamefont {J.~R.}\ \bibnamefont
  {Beamish}}, \bibinfo {author} {\bibfnamefont {A.~D.}\ \bibnamefont
  {Fefferman}}, \bibinfo {author} {\bibfnamefont {A.}~\bibnamefont {Haziot}},
  \bibinfo {author} {\bibfnamefont {X.}~\bibnamefont {Rojas}}, \ and\ \bibinfo
  {author} {\bibfnamefont {S.}~\bibnamefont {Balibar}},\ }\href@noop {}
  {\bibfield  {journal} {\bibinfo  {journal} {Phys. Rev. B}\ }\textbf {\bibinfo
  {volume} {85}},\ \bibinfo {pages} {180501} (\bibinfo {year}
  {2012})}\BibitemShut {NoStop}%
\bibitem [{\citenamefont {Haziot}\ \emph
  {et~al.}(2013{\natexlab{a}})\citenamefont {Haziot}, \citenamefont
  {Fefferman}, \citenamefont {Beamish},\ and\ \citenamefont
  {Balibar}}]{Haziot_etal_PRB_2013}%
  \BibitemOpen
  \bibfield  {author} {\bibinfo {author} {\bibfnamefont {A.}~\bibnamefont
  {Haziot}}, \bibinfo {author} {\bibfnamefont {A.~D.}\ \bibnamefont
  {Fefferman}}, \bibinfo {author} {\bibfnamefont {J.~R.}\ \bibnamefont
  {Beamish}}, \ and\ \bibinfo {author} {\bibfnamefont {S.}~\bibnamefont
  {Balibar}},\ }\href@noop {} {\bibfield  {journal} {\bibinfo  {journal} {Phys.
  Rev. B}\ }\textbf {\bibinfo {volume} {87}},\ \bibinfo {pages} {060509}
  (\bibinfo {year} {2013}{\natexlab{a}})}\BibitemShut {NoStop}%
\bibitem [{\citenamefont {Iwasa}(2013)}]{Iwasa_JLTP_2013}%
  \BibitemOpen
  \bibfield  {author} {\bibinfo {author} {\bibfnamefont {I.}~\bibnamefont
  {Iwasa}},\ }\href@noop {} {\bibfield  {journal} {\bibinfo  {journal} {J. Low
  Temp. Phys.}\ }\textbf {\bibinfo {volume} {171}},\ \bibinfo {pages} {30}
  (\bibinfo {year} {2013})}\BibitemShut {NoStop}%
\bibitem [{\citenamefont {Choi}\ \emph {et~al.}(2010)\citenamefont {Choi},
  \citenamefont {Takahashi}, \citenamefont {Kono},\ and\ \citenamefont
  {Kim}}]{Choi_and_Takahashi_etal_Science_2010}%
  \BibitemOpen
  \bibfield  {author} {\bibinfo {author} {\bibfnamefont {H.}~\bibnamefont
  {Choi}}, \bibinfo {author} {\bibfnamefont {D.}~\bibnamefont {Takahashi}},
  \bibinfo {author} {\bibfnamefont {K.}~\bibnamefont {Kono}}, \ and\ \bibinfo
  {author} {\bibfnamefont {E.}~\bibnamefont {Kim}},\ }\href@noop {} {\bibfield
  {journal} {\bibinfo  {journal} {Science}\ }\textbf {\bibinfo {volume}
  {330}},\ \bibinfo {pages} {1512} (\bibinfo {year} {2010})}\BibitemShut
  {NoStop}%
\bibitem [{\citenamefont {Choi}\ \emph
  {et~al.}(2012{\natexlab{a}})\citenamefont {Choi}, \citenamefont {Takahashi},
  \citenamefont {Choi}, \citenamefont {Kono},\ and\ \citenamefont
  {Kim}}]{Choi_and_Takahashi_etal_PRL_2012}%
  \BibitemOpen
  \bibfield  {author} {\bibinfo {author} {\bibfnamefont {H.}~\bibnamefont
  {Choi}}, \bibinfo {author} {\bibfnamefont {D.}~\bibnamefont {Takahashi}},
  \bibinfo {author} {\bibfnamefont {W.}~\bibnamefont {Choi}}, \bibinfo {author}
  {\bibfnamefont {K.}~\bibnamefont {Kono}}, \ and\ \bibinfo {author}
  {\bibfnamefont {E.}~\bibnamefont {Kim}},\ }\href@noop {} {\bibfield
  {journal} {\bibinfo  {journal} {Phys. Rev. Lett.}\ }\textbf {\bibinfo
  {volume} {108}},\ \bibinfo {pages} {105302} (\bibinfo {year}
  {2012}{\natexlab{a}})}\BibitemShut {NoStop}%
\bibitem [{\citenamefont {Choi}\ \emph
  {et~al.}(2012{\natexlab{b}})\citenamefont {Choi}, \citenamefont {Takahashi},
  \citenamefont {Kim}, \citenamefont {Choi}, \citenamefont {Kono},\ and\
  \citenamefont {Kim}}]{Choi_etal_PRB_2012}%
  \BibitemOpen
  \bibfield  {author} {\bibinfo {author} {\bibfnamefont {W.}~\bibnamefont
  {Choi}}, \bibinfo {author} {\bibfnamefont {D.}~\bibnamefont {Takahashi}},
  \bibinfo {author} {\bibfnamefont {D.~Y.}\ \bibnamefont {Kim}}, \bibinfo
  {author} {\bibfnamefont {H.}~\bibnamefont {Choi}}, \bibinfo {author}
  {\bibfnamefont {K.}~\bibnamefont {Kono}}, \ and\ \bibinfo {author}
  {\bibfnamefont {E.}~\bibnamefont {Kim}},\ }\href@noop {} {\bibfield
  {journal} {\bibinfo  {journal} {Phys. Rev. B}\ }\textbf {\bibinfo {volume}
  {86}},\ \bibinfo {pages} {174505} (\bibinfo {year}
  {2012}{\natexlab{b}})}\BibitemShut {NoStop}%
\bibitem [{\citenamefont {Suzuki}(1973)}]{suzuki1973plastic}%
  \BibitemOpen
  \bibfield  {author} {\bibinfo {author} {\bibfnamefont {H.}~\bibnamefont
  {Suzuki}},\ }\href@noop {} {\bibfield  {journal} {\bibinfo  {journal} {J.
  Phys. Soc. Jpn.}\ }\textbf {\bibinfo {volume} {35}},\ \bibinfo {pages} {1472}
  (\bibinfo {year} {1973})}\BibitemShut {NoStop}%
\bibitem [{\citenamefont {Wanner}\ \emph {et~al.}(1976)\citenamefont {Wanner},
  \citenamefont {Iwasa},\ and\ \citenamefont {Wales}}]{wanner1976evidence}%
  \BibitemOpen
  \bibfield  {author} {\bibinfo {author} {\bibfnamefont {R.}~\bibnamefont
  {Wanner}}, \bibinfo {author} {\bibfnamefont {I.}~\bibnamefont {Iwasa}}, \
  and\ \bibinfo {author} {\bibfnamefont {S.}~\bibnamefont {Wales}},\
  }\href@noop {} {\bibfield  {journal} {\bibinfo  {journal} {Solid State
  Comm.}\ }\textbf {\bibinfo {volume} {18}},\ \bibinfo {pages} {853} (\bibinfo
  {year} {1976})}\BibitemShut {NoStop}%
\bibitem [{\citenamefont {Iwasa}\ and\ \citenamefont
  {Suzuki}(1980)}]{iwasa1980sound}%
  \BibitemOpen
  \bibfield  {author} {\bibinfo {author} {\bibfnamefont {I.}~\bibnamefont
  {Iwasa}}\ and\ \bibinfo {author} {\bibfnamefont {H.}~\bibnamefont {Suzuki}},\
  }\href@noop {} {\bibfield  {journal} {\bibinfo  {journal} {J. Phys. Soc.
  Jpn.}\ }\textbf {\bibinfo {volume} {49}},\ \bibinfo {pages} {1722} (\bibinfo
  {year} {1980})}\BibitemShut {NoStop}%
\bibitem [{\citenamefont {Iwasa}\ \emph {et~al.}(1987)\citenamefont {Iwasa},
  \citenamefont {Suzuki}, \citenamefont {Nakajima}, \citenamefont {Suzuki},
  \citenamefont {Ando}, \citenamefont {Yonenaga}, \citenamefont {Takebe},\ and\
  \citenamefont {Sumino}}]{iwasa1987observation}%
  \BibitemOpen
  \bibfield  {author} {\bibinfo {author} {\bibfnamefont {I.}~\bibnamefont
  {Iwasa}}, \bibinfo {author} {\bibfnamefont {H.}~\bibnamefont {Suzuki}},
  \bibinfo {author} {\bibfnamefont {T.}~\bibnamefont {Nakajima}}, \bibinfo
  {author} {\bibfnamefont {S.}~\bibnamefont {Suzuki}}, \bibinfo {author}
  {\bibfnamefont {M.}~\bibnamefont {Ando}}, \bibinfo {author} {\bibfnamefont
  {I.}~\bibnamefont {Yonenaga}}, \bibinfo {author} {\bibfnamefont
  {M.}~\bibnamefont {Takebe}}, \ and\ \bibinfo {author} {\bibfnamefont
  {K.}~\bibnamefont {Sumino}},\ }\href@noop {} {\bibfield  {journal} {\bibinfo
  {journal} {J. Phys. Soc. Jpn.}\ }\textbf {\bibinfo {volume} {56}},\ \bibinfo
  {pages} {4225} (\bibinfo {year} {1987})}\BibitemShut {NoStop}%
\bibitem [{\citenamefont {Granato}\ and\ \citenamefont
  {L{\"u}cke}(1956)}]{granato1956theory}%
  \BibitemOpen
  \bibfield  {author} {\bibinfo {author} {\bibfnamefont {A.~V.}\ \bibnamefont
  {Granato}}\ and\ \bibinfo {author} {\bibfnamefont {K.}~\bibnamefont
  {L{\"u}cke}},\ }\href@noop {} {\bibfield  {journal} {\bibinfo  {journal} {J.
  Appl. Phys.}\ }\textbf {\bibinfo {volume} {27}},\ \bibinfo {pages} {583}
  (\bibinfo {year} {1956})}\BibitemShut {NoStop}%
\bibitem [{\citenamefont {Zhou}\ \emph {et~al.}(2012)\citenamefont {Zhou},
  \citenamefont {Su}, \citenamefont {Graf}, \citenamefont {Reichhardt},
  \citenamefont {Balatsky},\ and\ \citenamefont
  {Beyerlein}}]{zhou2012dislocation}%
  \BibitemOpen
  \bibfield  {author} {\bibinfo {author} {\bibfnamefont {C.}~\bibnamefont
  {Zhou}}, \bibinfo {author} {\bibfnamefont {J.-j.}\ \bibnamefont {Su}},
  \bibinfo {author} {\bibfnamefont {M.~J.}\ \bibnamefont {Graf}}, \bibinfo
  {author} {\bibfnamefont {C.}~\bibnamefont {Reichhardt}}, \bibinfo {author}
  {\bibfnamefont {A.~V.}\ \bibnamefont {Balatsky}}, \ and\ \bibinfo {author}
  {\bibfnamefont {I.~J.}\ \bibnamefont {Beyerlein}},\ }\href@noop {} {\bibfield
   {journal} {\bibinfo  {journal} {Phil. Mag. Lett.}\ }\textbf {\bibinfo
  {volume} {92}},\ \bibinfo {pages} {608} (\bibinfo {year} {2012})}\BibitemShut
  {NoStop}%
\bibitem [{\citenamefont {Haziot}\ \emph
  {et~al.}(2013{\natexlab{b}})\citenamefont {Haziot}, \citenamefont {Rojas},
  \citenamefont {Fefferman}, \citenamefont {Beamish},\ and\ \citenamefont
  {Balibar}}]{haziot2013giant}%
  \BibitemOpen
  \bibfield  {author} {\bibinfo {author} {\bibfnamefont {A.}~\bibnamefont
  {Haziot}}, \bibinfo {author} {\bibfnamefont {X.}~\bibnamefont {Rojas}},
  \bibinfo {author} {\bibfnamefont {A.~D.}\ \bibnamefont {Fefferman}}, \bibinfo
  {author} {\bibfnamefont {J.~R.}\ \bibnamefont {Beamish}}, \ and\ \bibinfo
  {author} {\bibfnamefont {S.}~\bibnamefont {Balibar}},\ }\href@noop {}
  {\bibfield  {journal} {\bibinfo  {journal} {Phys. Rev. Lett.}\ }\textbf
  {\bibinfo {volume} {110}},\ \bibinfo {pages} {035301} (\bibinfo {year}
  {2013}{\natexlab{b}})}\BibitemShut {NoStop}%
\bibitem [{\citenamefont {Fefferman}\ \emph {et~al.}(2014)\citenamefont
  {Fefferman}, \citenamefont {Souris}, \citenamefont {Haziot}, \citenamefont
  {Beamish},\ and\ \citenamefont {Balibar}}]{fefferman2014dislocation}%
  \BibitemOpen
  \bibfield  {author} {\bibinfo {author} {\bibfnamefont {A.}~\bibnamefont
  {Fefferman}}, \bibinfo {author} {\bibfnamefont {F.}~\bibnamefont {Souris}},
  \bibinfo {author} {\bibfnamefont {A.}~\bibnamefont {Haziot}}, \bibinfo
  {author} {\bibfnamefont {J.}~\bibnamefont {Beamish}}, \ and\ \bibinfo
  {author} {\bibfnamefont {S.}~\bibnamefont {Balibar}},\ }\href@noop {}
  {\bibfield  {journal} {\bibinfo  {journal} {Phys. Rev. B}\ }\textbf {\bibinfo
  {volume} {89}},\ \bibinfo {pages} {014105} (\bibinfo {year}
  {2014})}\BibitemShut {NoStop}%
\bibitem [{\citenamefont {Greenberg}\ \emph {et~al.}(1971)\citenamefont
  {Greenberg}, \citenamefont {Thomlinson},\ and\ \citenamefont
  {Richardson}}]{greenberg1971isotopic}%
  \BibitemOpen
  \bibfield  {author} {\bibinfo {author} {\bibfnamefont {A.~S.}\ \bibnamefont
  {Greenberg}}, \bibinfo {author} {\bibfnamefont {W.~C.}\ \bibnamefont
  {Thomlinson}}, \ and\ \bibinfo {author} {\bibfnamefont {R.~C.}\ \bibnamefont
  {Richardson}},\ }\href@noop {} {\bibfield  {journal} {\bibinfo  {journal}
  {Phys. Rev. Lett.}\ }\textbf {\bibinfo {volume} {27}},\ \bibinfo {pages}
  {179} (\bibinfo {year} {1971})}\BibitemShut {NoStop}%
\bibitem [{\citenamefont {Kim}\ and\ \citenamefont
  {Chan}(2006)}]{kim2006supersolid}%
  \BibitemOpen
  \bibfield  {author} {\bibinfo {author} {\bibfnamefont {E.}~\bibnamefont
  {Kim}}\ and\ \bibinfo {author} {\bibfnamefont {M.~H.~W.}\ \bibnamefont
  {Chan}},\ }\href@noop {} {\bibfield  {journal} {\bibinfo  {journal} {Phys.
  Rev. Lett.}\ }\textbf {\bibinfo {volume} {97}},\ \bibinfo {pages} {115302}
  (\bibinfo {year} {2006})}\BibitemShut {NoStop}%
\bibitem [{\citenamefont {Pushkarov}(2001)}]{pushkarov2001quasiparticle}%
  \BibitemOpen
  \bibfield  {author} {\bibinfo {author} {\bibfnamefont {D.~I.}\ \bibnamefont
  {Pushkarov}},\ }\href@noop {} {\bibfield  {journal} {\bibinfo  {journal}
  {Phys. Rep.}\ }\textbf {\bibinfo {volume} {354}},\ \bibinfo {pages} {411}
  (\bibinfo {year} {2001})}\BibitemShut {NoStop}%
\bibitem [{\citenamefont {Pushkarov}(2012)}]{pushkarov2012vacancy}%
  \BibitemOpen
  \bibfield  {author} {\bibinfo {author} {\bibfnamefont {D.~I.}\ \bibnamefont
  {Pushkarov}},\ }\href@noop {} {\bibfield  {journal} {\bibinfo  {journal}
  {Europhys. Lett.}\ }\textbf {\bibinfo {volume} {97}},\ \bibinfo {pages}
  {56002} (\bibinfo {year} {2012})}\BibitemShut {NoStop}%
\bibitem [{\citenamefont {Yagi}\ \emph {et~al.}(2011)\citenamefont {Yagi},
  \citenamefont {Kitamura}, \citenamefont {Shimizu}, \citenamefont {Yasuta},\
  and\ \citenamefont {Kubota}}]{yagi2011probable}%
  \BibitemOpen
  \bibfield  {author} {\bibinfo {author} {\bibfnamefont {M.}~\bibnamefont
  {Yagi}}, \bibinfo {author} {\bibfnamefont {A.}~\bibnamefont {Kitamura}},
  \bibinfo {author} {\bibfnamefont {N.}~\bibnamefont {Shimizu}}, \bibinfo
  {author} {\bibfnamefont {Y.}~\bibnamefont {Yasuta}}, \ and\ \bibinfo {author}
  {\bibfnamefont {M.}~\bibnamefont {Kubota}},\ }\href@noop {} {\bibfield
  {journal} {\bibinfo  {journal} {J. Low Temp. Phys.}\ }\textbf {\bibinfo
  {volume} {162}},\ \bibinfo {pages} {492} (\bibinfo {year}
  {2011})}\BibitemShut {NoStop}%
\bibitem [{\citenamefont {Fear}\ \emph {et~al.}(2016)\citenamefont {Fear},
  \citenamefont {Walmsley}, \citenamefont {Zmeev}, \citenamefont
  {M{\"a}kinen},\ and\ \citenamefont {Golov}}]{fear2016no}%
  \BibitemOpen
  \bibfield  {author} {\bibinfo {author} {\bibfnamefont {M.~J.}\ \bibnamefont
  {Fear}}, \bibinfo {author} {\bibfnamefont {P.~M.}\ \bibnamefont {Walmsley}},
  \bibinfo {author} {\bibfnamefont {D.~E.}\ \bibnamefont {Zmeev}}, \bibinfo
  {author} {\bibfnamefont {J.~T.}\ \bibnamefont {M{\"a}kinen}}, \ and\ \bibinfo
  {author} {\bibfnamefont {A.~I.}\ \bibnamefont {Golov}},\ }\href@noop {}
  {\bibfield  {journal} {\bibinfo  {journal} {J. Low Temp. Phys.}\ }\textbf
  {\bibinfo {volume} {183}},\ \bibinfo {pages} {106} (\bibinfo {year}
  {2016})}\BibitemShut {NoStop}%
\bibitem [{\citenamefont {Choi}\ \emph {et~al.}(2017)\citenamefont {Choi},
  \citenamefont {Tsuiki}, \citenamefont {Takahashi}, \citenamefont {Choi},
  \citenamefont {Kono}, \citenamefont {Shirahama},\ and\ \citenamefont
  {Kim}}]{choi2017superfluid}%
  \BibitemOpen
  \bibfield  {author} {\bibinfo {author} {\bibfnamefont {J.}~\bibnamefont
  {Choi}}, \bibinfo {author} {\bibfnamefont {T.}~\bibnamefont {Tsuiki}},
  \bibinfo {author} {\bibfnamefont {D.}~\bibnamefont {Takahashi}}, \bibinfo
  {author} {\bibfnamefont {H.}~\bibnamefont {Choi}}, \bibinfo {author}
  {\bibfnamefont {K.}~\bibnamefont {Kono}}, \bibinfo {author} {\bibfnamefont
  {K.}~\bibnamefont {Shirahama}}, \ and\ \bibinfo {author} {\bibfnamefont
  {E.}~\bibnamefont {Kim}},\ }\href@noop {} {\bibfield  {journal} {\bibinfo
  {journal} {arXiv:1701.07190}\ } (\bibinfo {year} {2017})}\BibitemShut
  {NoStop}%
\bibitem [{\citenamefont {Choi}\ \emph {et~al.}(2015)\citenamefont {Choi},
  \citenamefont {Shin},\ and\ \citenamefont {Kim}}]{choi2015frequency}%
  \BibitemOpen
  \bibfield  {author} {\bibinfo {author} {\bibfnamefont {J.}~\bibnamefont
  {Choi}}, \bibinfo {author} {\bibfnamefont {J.}~\bibnamefont {Shin}}, \ and\
  \bibinfo {author} {\bibfnamefont {E.}~\bibnamefont {Kim}},\ }\href@noop {}
  {\bibfield  {journal} {\bibinfo  {journal} {Phys. Rev. B}\ }\textbf {\bibinfo
  {volume} {92}},\ \bibinfo {pages} {144505} (\bibinfo {year}
  {2015})}\BibitemShut {NoStop}%
\bibitem [{\citenamefont {Tsuiki}\ \emph {et~al.}()\citenamefont {Tsuiki},
  \citenamefont {Takahashi}, \citenamefont {Okuda}, \citenamefont {Kono},\ and\
  \citenamefont {Shirahama}}]{TsuikiPreparation}%
  \BibitemOpen
  \bibfield  {author} {\bibinfo {author} {\bibfnamefont {T.}~\bibnamefont
  {Tsuiki}}, \bibinfo {author} {\bibfnamefont {D.}~\bibnamefont {Takahashi}},
  \bibinfo {author} {\bibfnamefont {Y.}~\bibnamefont {Okuda}}, \bibinfo
  {author} {\bibfnamefont {K.}~\bibnamefont {Kono}}, \ and\ \bibinfo {author}
  {\bibfnamefont {K.}~\bibnamefont {Shirahama}},\ }\href@noop {} {}\bibinfo
  {note} {(in preparation)}\BibitemShut {NoStop}%
\bibitem [{\citenamefont {Reppy}\ \emph {et~al.}(2012)\citenamefont {Reppy},
  \citenamefont {Mi}, \citenamefont {Justin},\ and\ \citenamefont
  {Mueller}}]{reppy2012interpreting}%
  \BibitemOpen
  \bibfield  {author} {\bibinfo {author} {\bibfnamefont {J.~D.}\ \bibnamefont
  {Reppy}}, \bibinfo {author} {\bibfnamefont {X.}~\bibnamefont {Mi}}, \bibinfo
  {author} {\bibfnamefont {A.}~\bibnamefont {Justin}}, \ and\ \bibinfo {author}
  {\bibfnamefont {E.~J.}\ \bibnamefont {Mueller}},\ }\href@noop {} {\bibfield
  {journal} {\bibinfo  {journal} {J. Low Temp. Phys.}\ }\textbf {\bibinfo
  {volume} {168}},\ \bibinfo {pages} {175} (\bibinfo {year}
  {2012})}\BibitemShut {NoStop}%
\bibitem [{\citenamefont {Takahashi}\ and\ \citenamefont
  {Kono}(2006)}]{takahashi2006new}%
  \BibitemOpen
  \bibfield  {author} {\bibinfo {author} {\bibfnamefont {D.}~\bibnamefont
  {Takahashi}}\ and\ \bibinfo {author} {\bibfnamefont {K.}~\bibnamefont
  {Kono}},\ }in\ \href@noop {} {\emph {\bibinfo {booktitle} {AIP Conference
  Proceedings}}},\ Vol.\ \bibinfo {volume} {850}\ (\bibinfo {organization}
  {AIP},\ \bibinfo {year} {2006})\ pp.\ \bibinfo {pages}
  {1567--1568}\BibitemShut {NoStop}%
\bibitem [{FUJ()}]{FUJI}%
  \BibitemOpen
  \href@noop {} {}\bibinfo {howpublished}
  {{http://www.fujicera.co.jp/index\_e.html}}\BibitemShut {NoStop}%
\bibitem [{\citenamefont {Kim}\ \emph {et~al.}(2011)\citenamefont {Kim},
  \citenamefont {Choi}, \citenamefont {Choi}, \citenamefont {Kwon},
  \citenamefont {Kim},\ and\ \citenamefont {Kim}}]{DYKim_etal_PRB_2011}%
  \BibitemOpen
  \bibfield  {author} {\bibinfo {author} {\bibfnamefont {D.~Y.}\ \bibnamefont
  {Kim}}, \bibinfo {author} {\bibfnamefont {H.}~\bibnamefont {Choi}}, \bibinfo
  {author} {\bibfnamefont {W.}~\bibnamefont {Choi}}, \bibinfo {author}
  {\bibfnamefont {S.}~\bibnamefont {Kwon}}, \bibinfo {author} {\bibfnamefont
  {E.}~\bibnamefont {Kim}}, \ and\ \bibinfo {author} {\bibfnamefont {H.~C.}\
  \bibnamefont {Kim}},\ }\href@noop {} {\bibfield  {journal} {\bibinfo
  {journal} {Phys. Rev. B}\ }\textbf {\bibinfo {volume} {83}},\ \bibinfo
  {pages} {052503} (\bibinfo {year} {2011})}\BibitemShut {NoStop}%
\bibitem [{\citenamefont {Iwasa}()}]{Iwasa_pc}%
  \BibitemOpen
  \bibfield  {author} {\bibinfo {author} {\bibfnamefont {I.}~\bibnamefont
  {Iwasa}},\ }\href@noop {} {}\bibinfo {note} {(private
  communication)}\BibitemShut {NoStop}%
\bibitem [{\citenamefont {Huan}\ \emph {et~al.}(2016)\citenamefont {Huan},
  \citenamefont {Kim}, \citenamefont {Candela},\ and\ \citenamefont
  {Sullivan}}]{huan2016quantum}%
  \BibitemOpen
  \bibfield  {author} {\bibinfo {author} {\bibfnamefont {C.}~\bibnamefont
  {Huan}}, \bibinfo {author} {\bibfnamefont {S.~S.}\ \bibnamefont {Kim}},
  \bibinfo {author} {\bibfnamefont {D.}~\bibnamefont {Candela}}, \ and\
  \bibinfo {author} {\bibfnamefont {N.~S.}\ \bibnamefont {Sullivan}},\
  }\href@noop {} {\bibfield  {journal} {\bibinfo  {journal} {J. Low Temp.
  Phys.}\ }\textbf {\bibinfo {volume} {185}},\ \bibinfo {pages} {354} (\bibinfo
  {year} {2016})}\BibitemShut {NoStop}%
\bibitem [{\citenamefont {Greenberg}\ \emph {et~al.}(1972)\citenamefont
  {Greenberg}, \citenamefont {Thomlinson},\ and\ \citenamefont
  {Richardson}}]{greenberg1972nuclear}%
  \BibitemOpen
  \bibfield  {author} {\bibinfo {author} {\bibfnamefont {A.~S.}\ \bibnamefont
  {Greenberg}}, \bibinfo {author} {\bibfnamefont {W.~C.}\ \bibnamefont
  {Thomlinson}}, \ and\ \bibinfo {author} {\bibfnamefont {R.~C.}\ \bibnamefont
  {Richardson}},\ }\href@noop {} {\bibfield  {journal} {\bibinfo  {journal} {J.
  Low Temp. Phys.}\ }\textbf {\bibinfo {volume} {8}},\ \bibinfo {pages} {3}
  (\bibinfo {year} {1972})}\BibitemShut {NoStop}%
\bibitem [{\citenamefont {Shah}(1999)}]{shah_thesis_simmons}%
  \BibitemOpen
  \bibfield  {author} {\bibinfo {author} {\bibfnamefont {R.~S.}\ \bibnamefont
  {Shah}},\ }\emph {\bibinfo {title} {X-ray and pressure measurements of helium
  solids}},\ \href@noop {} {Ph.D. thesis},\ \bibinfo  {school} {Univ. Illinois
  at Urbana-Champaign} (\bibinfo {year} {1999})\BibitemShut {NoStop}%
\bibitem [{\citenamefont {Edwards}\ and\ \citenamefont
  {Pandorf}(1965)}]{edwards1965heat}%
  \BibitemOpen
  \bibfield  {author} {\bibinfo {author} {\bibfnamefont {D.~O.}\ \bibnamefont
  {Edwards}}\ and\ \bibinfo {author} {\bibfnamefont {R.~C.}\ \bibnamefont
  {Pandorf}},\ }\href@noop {} {\bibfield  {journal} {\bibinfo  {journal} {Phys.
  Rev.}\ }\textbf {\bibinfo {volume} {140}},\ \bibinfo {pages} {A816} (\bibinfo
  {year} {1965})}\BibitemShut {NoStop}%
\bibitem [{\citenamefont {Kang}\ \emph {et~al.}(2015)\citenamefont {Kang},
  \citenamefont {Yoon},\ and\ \citenamefont {Kim}}]{kang2015modified}%
  \BibitemOpen
  \bibfield  {author} {\bibinfo {author} {\bibfnamefont {E.~S.}\ \bibnamefont
  {Kang}}, \bibinfo {author} {\bibfnamefont {H.}~\bibnamefont {Yoon}}, \ and\
  \bibinfo {author} {\bibfnamefont {E.}~\bibnamefont {Kim}},\ }\href@noop {}
  {\bibfield  {journal} {\bibinfo  {journal} {J. Phys. Soc. Jpn.}\ }\textbf
  {\bibinfo {volume} {84}},\ \bibinfo {pages} {034602} (\bibinfo {year}
  {2015})}\BibitemShut {NoStop}%
\bibitem [{\citenamefont {Iwasa}\ \emph {et~al.}(1979)\citenamefont {Iwasa},
  \citenamefont {Araki},\ and\ \citenamefont {Suzuki}}]{iwasa1979temperature}%
  \BibitemOpen
  \bibfield  {author} {\bibinfo {author} {\bibfnamefont {I.}~\bibnamefont
  {Iwasa}}, \bibinfo {author} {\bibfnamefont {K.}~\bibnamefont {Araki}}, \ and\
  \bibinfo {author} {\bibfnamefont {H.}~\bibnamefont {Suzuki}},\ }\href@noop {}
  {\bibfield  {journal} {\bibinfo  {journal} {J. Phys. Soc. Jpn.}\ }\textbf
  {\bibinfo {volume} {46}},\ \bibinfo {pages} {1119} (\bibinfo {year}
  {1979})}\BibitemShut {NoStop}%
\bibitem [{\citenamefont {Kim}\ and\ \citenamefont
  {Chan}(2014)}]{kim2014upper}%
  \BibitemOpen
  \bibfield  {author} {\bibinfo {author} {\bibfnamefont {D.~Y.}\ \bibnamefont
  {Kim}}\ and\ \bibinfo {author} {\bibfnamefont {M.~H.~W.}\ \bibnamefont
  {Chan}},\ }\href@noop {} {\bibfield  {journal} {\bibinfo  {journal} {Phys.
  Rev. B}\ }\textbf {\bibinfo {volume} {90}},\ \bibinfo {pages} {064503}
  (\bibinfo {year} {2014})}\BibitemShut {NoStop}%
\bibitem [{\citenamefont {Maris}(2012)}]{maris2012effect}%
  \BibitemOpen
  \bibfield  {author} {\bibinfo {author} {\bibfnamefont {H.~J.}\ \bibnamefont
  {Maris}},\ }\href@noop {} {\bibfield  {journal} {\bibinfo  {journal} {Phys.
  Rev. B}\ }\textbf {\bibinfo {volume} {86}},\ \bibinfo {pages} {020502}
  (\bibinfo {year} {2012})}\BibitemShut {NoStop}%
\bibitem [{\citenamefont {Kim}\ and\ \citenamefont
  {Chan}(2012)}]{kim2012absence}%
  \BibitemOpen
  \bibfield  {author} {\bibinfo {author} {\bibfnamefont {D.~Y.}\ \bibnamefont
  {Kim}}\ and\ \bibinfo {author} {\bibfnamefont {M.~H.~W.}\ \bibnamefont
  {Chan}},\ }\href@noop {} {\bibfield  {journal} {\bibinfo  {journal} {Phys.
  Rev. Lett.}\ }\textbf {\bibinfo {volume} {109}},\ \bibinfo {pages} {155301}
  (\bibinfo {year} {2012})}\BibitemShut {NoStop}%
\bibitem [{\citenamefont {Dattoli}\ and\ \citenamefont
  {Quattromini}(2010)}]{dattoli2010note}%
  \BibitemOpen
  \bibfield  {author} {\bibinfo {author} {\bibfnamefont {G.}~\bibnamefont
  {Dattoli}}\ and\ \bibinfo {author} {\bibfnamefont {M.}~\bibnamefont
  {Quattromini}},\ }\href@noop {} {\bibfield  {journal} {\bibinfo  {journal}
  {arXiv:1009.3788}\ } (\bibinfo {year} {2010})}\BibitemShut {NoStop}%
\bibitem [{\citenamefont {Cooper}(2008)}]{cooper2008rapidly}%
  \BibitemOpen
  \bibfield  {author} {\bibinfo {author} {\bibfnamefont {N.~R.}\ \bibnamefont
  {Cooper}},\ }\href@noop {} {\bibfield  {journal} {\bibinfo  {journal} {Adv.
  Phys.}\ }\textbf {\bibinfo {volume} {57}},\ \bibinfo {pages} {539} (\bibinfo
  {year} {2008})}\BibitemShut {NoStop}%
\end{thebibliography}
%


\end{document}